\documentclass[12pt]{iopart}
\usepackage[utf8]{inputenc}
\usepackage[T1]{fontenc}
\usepackage[english]{babel}
\usepackage{color}
\expandafter\let\csname equation*\endcsname\relax
\expandafter\let\csname endequation*\endcsname\relax
\usepackage{empheq}
\usepackage{graphicx}
\usepackage[doublespacing]{setspace}
\usepackage[titletoc,toc,title]{appendix}
\usepackage[left=2cm,right=2cm,top=2cm,bottom=2cm]{geometry}

\usepackage{hyperref}
\usepackage[numbers]{natbib}
\usepackage{har2nat}
\hypersetup{breaklinks=true,colorlinks=true,linkcolor=blue,citecolor=blue,filecolor=magenta,urlcolor=cyan}

\renewcommand{\harvardurl}[1]{\url{#1}}

\usepackage{orcidlink}

\usepackage{textcomp}

\newcommand{\citemany}[2]{\protect\cite{#1}--\protect\cite{#2}}

\setcitestyle{square}

\begin{document}

\title[Nuclear-DFT electromagnetic moments]{Electromagnetic moments in the Sn-Gd region determined within nuclear DFT}
\author{H Wibowo\orcidlink{0000-0003-4093-0600}$^1$, B C  Backes\orcidlink{0000-0002-4490-2802}$^1$, J Dobaczewski\orcidlink{0000-0002-4158-3770}$^{1,2}$, R P de Groote\orcidlink{0000-0003-4942-1220}$^{3,4}$, A Nagpal\orcidlink{0000-0001-9206-712X}$^1$, A Sánchez-Fernández\orcidlink{0000-0003-3502-6668}$^{1,5,6}$, X Sun\orcidlink{0000-0002-0130-6269}$^1$, J L Wood\orcidlink{0009-0005-5832-8345}$^7$}
\address{
$^1$ School of Physics, Engineering and Technology, University of York, Heslington, York YO10 5DD, United Kingdom \\
$^2$ Institute of Theoretical Physics, Faculty of Physics, University of Warsaw, ul. Pasteura 5, PL-02-093 Warsaw, Poland \\
$^3$ Accelerator Laboratory, Department of Physics, University of Jyv\"askyl\"a, PB 35(YFL) FIN-40351 Jyv\"askyl\"a, Finland \\
$^4$ KU Leuven, Instituut voor Kern-en Stralingsfysica, B-3001 Leuven, Belgium \\
$^5$ Institut d’Astronomie et d’Astrophysique, Université Libre de Bruxelles, Brussels, Belgium \\
$^6$ Brussels Laboratory of the Universe - BLU-ULB \\
$^7$ School of Physics, Georgia Institute of Technology, Atlanta, Georgia 30332-0430, USA}

\date{\today}

\ead{herlik.wibowo@york.ac.uk}

\begin{abstract}
Within the nuclear DFT framework, employing the Skyrme UNEDF1 functional and incorporating pairing correlations, we determined the spectroscopic electric quadrupole and magnetic dipole moments of the $\nu11/2^{-}$ and $\pi7/2^{+}$ configurations in heavy, deformed, open-shell odd nuclei with $50\leq Z \leq 64$. The notions of self-consistent shape and spin polarisations due to odd nucleons responsible for generating total electric quadrupole and magnetic dipole moments were transformed into detailed computational procedures. The alignment of intrinsic angular momentum along the axial symmetry axis, necessitating signature and time-reversal symmetry breaking, followed by the restoration of rotational symmetry, proved to be essential components of the method. In contrast, the restoration of particle number symmetry yields modifications of only about 1\%. With the isovector spin-spin terms of the functional previously adjusted in near doubly magic nuclei across the mass chart, the calculations were parameter-free. Effective charges and $g$-factors were not employed. A reasonably good agreement was achieved between the calculated and measured electric quadrupole moments. A similarly fair description of the magnetic dipole moments was obtained for the intruder configurations $\nu11/2^{-}$ alongside a poor description of those for $\pi7/2^{+}$. 
\end{abstract}

\submitto{\JPG}

\maketitle

\section{Introduction}

The advancement of laser spectroscopy techniques provides a wealth of information about the fundamental properties of nuclei, including nuclear spins, electromagnetic moments, and charge radii~\cite{(Yan23)}. Electric quadrupole and magnetic dipole moments are among the most commonly studied nuclear electromagnetic properties~\cite{Castel1990,Neyens2003}. The electric quadrupole moment serves as a crucial tool for investigating nuclear deformation, while the magnetic dipole moment offers insights into the configurations of valence nucleons (either particle or hole). For a fixed configuration, the single-particle estimate of the magnetic moment, known as the Schmidt limit~\cite{(Sch37)}, remains constant along the isotopic chain. Deviations of the measured magnetic moments from the Schmidt value can be significant indicators of changes in internal structure, core-polarisation effects, and nuclear two-body currents that influence the form of the magnetic dipole operator. Theoretical calculations of both quadrupole and magnetic dipole moments are often model-dependent; hence, comparisons with data can serve as benchmarks to assess not only the validity of nuclear models but also to enhance understanding of the underlying physics of complex systems like nuclei.

\begin{figure}[ht]
\begin{center}
\includegraphics[width=0.80\textwidth]{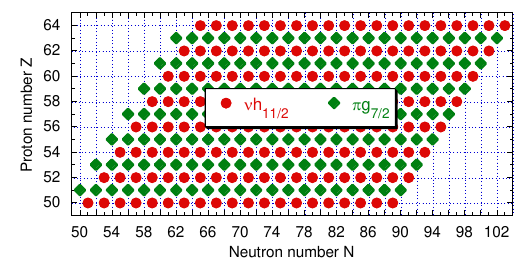}
\end{center}
\caption{\label{fig1} Schematic illustration of the set of odd-$Z$ (diamonds) and odd-$N$ (circles) nuclei considered in the present work.}
\end{figure}

In the present work, we expand the scope of the results presented in~\cite{(Bon23c)} to encompass the region of lighter elements, ranging from tin to gadolinium. Within nuclear density functional theory (DFT), we determine the spectroscopic electric quadrupole and magnetic dipole moments of the nuclei depicted in figure~\ref{fig1}. Our dual focus here is on studying (i) the neutron intruder states 1h$_{11/2}$ instead of 1i$_{13/2}$ and (ii) the proton spin-orbit partner states 1g$_{7/2}$ rather than the intruder states 1h$_{11/2}$. The goal is to identify and compare the principal features of nucleon polarisation effects for parallel and antiparallel coupling of spin and orbital angular momenta, respectively, and to investigate the properties of the selected orbitals upon crossing the magic neutron shell $N=82$. The abundance of experimental data, e.g., the new experimental data on the electric quadrupole and magnetic dipole moments in tin isotopes~\cite{Yordanov2020} and antimony isotopes~\cite{(Lec23)}, serves as an advantage of studying the neutron 1h$_{11/2}$ and proton 1g$_{7/2}$ configurations. 

Non-perturbative shape and angular-momentum polarisations are the main features that distinguish the nuclear DFT from approaches based on valence-space approximations. Within the DFT picture, the unpaired nucleon determines the properties of an odd nucleus in the following way. For small quadrupole deformations near the semi-magic systems, an $I>1/2$ hole (particle) in the magnetic sub-state having the largest angular-momentum projection $\Omega$ equal to the total angular momentum $I$, $\Omega=I$, induces the prolate (oblate) polarisation of the nucleus. This rule is an immediate consequence of the geometrically driven dependence of the orbital single-particle energy $\epsilon$ on the axial deformation $\beta$, that is, $\epsilon\sim\beta\left(3\Omega^2-I(I+1)\right)$.

Such dependence on $I$ and $\Omega$ originates from the expectation value of a small quadrupole perturbation, $\hat{H}_{1}\sim\beta\hat{Q}_{20}$, of the standard spherical mean-field eigenstates $|N\ell{}I\Omega\rangle$. From the Wigner-Eckart theorem, the expectation value $\langle{}N\ell{}I\Omega|\hat{Q}_{20}|N\ell{}I\Omega\rangle$ of the quadrupole perturbation is proportional to the Clebsch-Gordan coefficient $C^{I\Omega}_{I\Omega,20}=\left(3\Omega^2-I(I+1)\right)/\left((2I-1)I(I+1)(2I+3)\right)$~\cite{(Var88)}. Indeed, the orbital $\Omega=I$ then has the highest (lowest) energy $\epsilon\sim\beta{I}(2I-1)$ for $\beta>0$ ($\beta<0$) and thus determines the shape of the hole (particle) state.

In nuclear DFT, this perturbative picture undergoes an essential modification. Indeed, the non-zero quadrupole moment of a single hole (particle) occupying a magnetic sub-state with $\Omega=I$ induces a prolate (oblate) deformation to the total mean field, which, in turn, causes the non-zero quadrupole moments of the core nucleons. The induced total quadrupole moment of the core further deforms the entire mean field and, consequently, influences the quadrupole moment of the hole (particle). These mutual polarisations are summed up in the self-consistent solution, generating the odd nucleus's total quadrupole moment $Q$, which in open-shell systems becomes large, well beyond the perturbative effect. Since the mutual polarisations act throughout the full space of single-particle states, no effective polarisation charges are needed to determine $Q$.

Analogous to the self-consistent shape polarisations, the non-zero magnetic dipole moment of the odd nucleon induces non-zero magnetic dipole moments in the core nucleons, which, in turn, influence the magnetic dipole moment of the odd nucleon. In the self-consistent solution, these mutual angular-momentum polarisations (spin and orbital alike) produce the total magnetic dipole moment $\boldsymbol{\mu}$ of the odd nucleus in the same direction as the magnetic dipole moment of the odd nucleon. The time-odd mean-field sector of the energy functional determines the magnetic polarisation strength of the odd hole or particle. If the time-odd mean-field sector is neglected, only the magnetic moment of the odd nucleon survives. Again, since the mutual polarisations act throughout the full space of single-particle states, no effective $g$-factors are needed. As shown in~\cite{(Sas22c)}, this feature of nuclear DFT removes the adjustable effective parameters customarily used in valence-space approaches.

For axial shapes, and especially for determining the magnetic dipole moments $\boldsymbol{\mu}$, the nuclear DFT's polarisation picture requires the following two elements:
\begin{itemize}
    \item Aligning the odd particle's angular momentum along the axis of axial symmetry. Compared to other possible shape-alignment orientations, this allows for the largest values of the angular momentum projections and thus gives the strongest polarisation effects.
    \item Restoring the angular momentum symmetry, which allows for determining the spectroscopic moments that are directly comparable to experimental values.
\end{itemize}

In~\cite{(Sas22c),(Bon23c)}, we presented a brief literature overview of theoretical approaches used to determine nuclear electromagnetic moments; see also a recent review article~\cite{(Yan23)}.
 Since then, the work in this direction has continued in several publications, such as the
 shell model~\nocite{(Als23),(Ye23),(Pow22a),(Ish23),(Lec23),(Sah24)}\citemany{(Als23)}{(Sah24)},
 self-consistent Theory of Finite Fermi Systems (TFFS)~\cite{(Bor24),(Yue24)},
 Quasiparticle Phonon Nuclear Model (QPNM)~\cite{(Tab23)},
 interacting boson-fermion model with configuration mixing~\cite{(Gav22),(Gav23)},
 Hartree-Fock-Bogolyubov (HFB)~\cite{(Cub23),(Rys22a),(Web23)} and 
 Hartree-Fock plus BCS (HF+BCS) approaches~\cite{(Mor22)}, {\it ab initio} IMSRG with projected generator coordinate method (PGCM)~\cite{(Zho25)}, {\it ab initio} in-medium no-core shell model (IM-NSCM) with PGCM~\cite{(Fro22)}, 
 {\it ab initio} no-core shell-model~\cite{(Sar23),(Chu23)},
 {\it ab initio} valence-space in-medium similarity renormalisation group (VS-IMSRG)~\cite{(Pow22a),(Lec23),(Mul24)},
 covariant density-functional theory (CDFT)~\cite{(Ye23)},
 generator coordinate method within the multireference covariant density-functional theory (MR-CDFT)~\cite{(Zho24)},
 multi-reference energy density functional (MR-EDF) calculations~\cite{(Bal23)},
 and
 rigid triaxial rotor plus particle model~\cite{(Iva22)}.
In parallel, the methodology used in the present study was tested against experimental data in~\nocite{(deG22),(Ver22b),(Nie23),(Gra23),(deG24),(Kar24)}\citemany{(deG22)}{(Kar24)}.

In~\nocite{(Pow22a),(Ish23),(Lec23),(Sah24),(Bor24),(Yue24),(Tab23),(Gav22),(Gav23),(Cub23)}\citemany{(Pow22a)}{(Cub23)}, the results were obtained by employing effective charges and/or $g$-factors, with disparate values and justifications. Numerous diverse physical effects could be hidden and mixed under such numerical constants. These could include a limited size of the single-particle phase space, inadequate shape and/or angular-momentum polarisations, restrictions on conserved or broken symmetries or shape-alignment orientations, missing terms or deficiencies in the interactions or functionals, triaxial and/or octupole deformability, contributions from two-body meson-exchange currents, or maybe even a few more unknown unknowns, which potentially can all influence deviations between theory and experiment and deserve dedicated studies.

The paper is organised as follows. In section~\ref{Methodology}, we present details of the computational scheme employed in this work, and in section~\ref{Results}, we present the results obtained. Three subsections of section~\ref{Accuracy} discuss various aspects of the precision of the calculations related to the harmonic oscillator (HO) basis used, section~\ref{HO}, pairing strengths, section~\ref{pairing}, and effects of particle number projection (PNP), section~\ref{PNP}. 
Conclusions and outlook are presented in section~\ref{Conclusions|}.

\section{Methodology}\label{Methodology}

Following the concepts of shape and angular-momentum polarisations presented above, we outline the procedures we used to calculate nuclear electromagnetic moments systematically.
\begin{enumerate}
    \item In the first step, we performed the standard HF calculation for the spherical $^{100}\text{Sn}$ nucleus.
 
    \item In the next step, we broke the spherical and time-reversal symmetries of $^{100}\text{Sn}$ by introducing small axial quadrupole constraints $\bar{Q}_{20}$ of about $\pm 0.9$\,b, and a constraint on the $z$-component of the angular angular momentum ${J_z}$ for a small cranking frequency $\omega_{z}=0.001$\, MeV/$\hbar$. The constraint on $J_{z}$ (using a small cranking frequency $\omega_{z}$) enforces the alignment of the angular momenta along the axis of axial symmetry of the deformed nucleus. Together with the constraint on $J_{z}$, the two quadrupole constraints gave us weakly deformed prolate and oblate $^{100}\text{Sn}$ states with the projections of the single-particle angular momenta quantised along the axial-symmetry $z$-axis. In this way, in the spectrum of single-particle states, we could identify two Nilsson orbitals, $\nu[505]11/2^{-}$ and $\pi[404]7/2^{+}$, which originate from the spherical neutron 1h$_{11/2}$ and proton 1g$_{7/2}$ states, respectively.    
    The prolate and oblate single-particle states defined in this way were fixed and used to tag blocked quasiparticles for all open-shell odd nuclei considered in this work. In what follows, we refer to them as prolate and oblate tags.
    
    The tagging mechanism was implemented in the following way~\cite{(Dob09g)}. In each iteration of the HFB self-consistent procedure~\cite{(Rin80)}, the code scanned the two-component quasiparticle wave functions $\chi_\mu=\left(\begin{array}{c}B^*_\mu\\A^*_\mu\end{array}\right)$ and determined the overlaps ${\cal{O}}_{\mu}=\max\left((\phi_\nu|B^*_\mu),(\phi_\nu|\hat{T}^+A^*_{\mu}\hat{T})\right)$ between the fixed-tag single-particle wave function $\phi_\nu$ and the upper $B^*_\mu$ or time-reversed lower quasiparticle component $\hat{T}^+A^*_{\mu}\hat{T}$. Then, the maximum overlap ${\cal{O}}_{\mu}$ defined the quasiparticle state $\mu$ to be blocked. The tagging mechanism has the advantage of being insensitive to the energies of quasiparticles, which can change from one iteration to another. Most importantly, it is also insensitive to whether the quasiparticles are predominantly of the particle or hole character; that is, it allows for following them when they cross the Fermi energy.
    
    The tagging mechanism is based on the strong affinity between the self-consistent quasiparticle states and single-particle states near closed shells. This affinity is the crux of our method for following specific configurations across a range of particle numbers.

    In summary, we first defined the fixed single-particle states $\nu[505]11/2^{-}$ and $\pi[404]7/2^{+}$ by applying the quadrupole constraints $\bar{Q}_{20} = \pm 0.9$ b. Then, we selected the one-quasiparticle state by maximising the overlap $\mathcal{O}_{\mu}$ between the quasiparticle wave functions and the previously defined fixed single-particle states $\nu[505]11/2^{-}$ and $\pi[404]7/2^{+}$.
    
    \item In the next step, for each considered nucleus, depicted in figure~\ref{fig1}, we performed a tagged quasiparticle-blocking calculation by setting the average constant neutron and proton pairing gaps to 1\,MeV~\cite{(Dob04d)}. We also imposed axial quadrupole constraints for prolate and oblate tags with large values of $\bar{Q}_{20}=+10$ and $-10$\,b, respectively. This allowed us to determine stable starting reference solutions suitable for obtaining the final self-consistent states.

    \item In the next step, we determined the final self-consistent states by substituting the fixed pairing gaps with the neutron and proton pairing forces and relaxing the quadrupole-moment constraints. In this step, the pairing correlations were self-consistently determined. At the same time, during the self-consistent iteration, for both prolate and oblate tags, the quadrupole deformations slid back to lower quadrupole deformations, reaching the prolate and oblate minima, respectively. Both tags led to the same self-consistent solution in all cases where only a single minimum existed, whether prolate or oblate.

    \item In the last step, for all self-consistent solutions, we restored the rotational symmetry~\cite{Sheikh_2021} by projecting the intrinsic states on good angular momenta of $I=11/2$ or $7/2\,\hbar$ for the $\nu11/2^{-}$ and $\pi7/2^{+}$ configurations, respectively. At this point, the preserved axial symmetry and fixed quantised values of the angular momentum projections along the axial symmetry axis allowed for reducing the 3D integration over the three Euler angles~\cite{Sheikh_2021} to the 1D integration over the rotation about the axis perpendicular to the axial symmetry axis only. This reduces the CPU time required to perform the angular momentum projection (AMP) by about two orders of magnitude and makes it comparable to that needed to converge the HFB states. The spectroscopic electric quadrupole and magnetic dipole moments were then determined for symmetry-restored states.
\end{enumerate}

The methodology proposed for the first time in~\cite{(Bon23c)} and outlined in this section enables us to perform large-scale calculations of nuclear electromagnetic moments for many elements and long chains of isotopes, ranging from near-closed shells to open-shell configurations. The tagging mechanism allows us to follow given microscopic structures and configurations irrespective of their precise excitation energies and ordering in the spectra of odd nuclei. In this way, particular systematic agreement or disagreement patterns with data can be identified, and improvements in theoretical modelling can be proposed, implemented, studied, and tested.

\section{Results}\label{Results}

In this work, we employ the UNEDF1~\cite{(Kor12b)} parametrisation of the Skyrme functional and focus on the time-odd mean fields generated by the spin-spin interactions of the form $\boldsymbol{\sigma}_{1}\cdot\boldsymbol{\sigma}_{2}$. The standard Skyrme functionals, like UNEDF1, generate the time-odd mean fields via the following terms~\cite{(Dob95b),(Per04c)}: $C^{2}_{t}s^{2}_{t}(\textbf{r})$, $C^{\Delta s}_{t}\textbf{s}_{t}(\textbf{r})\cdot\nabla^{2}\textbf{s}_{t}(\textbf{r})$, $C^{T}_{t}\textbf{s}_{t}(\textbf{r})\cdot\textbf{T}_{t}(\textbf{r})$, $C^{j}_{t}j^{2}_{t}(\textbf{r})$, and $C^{\nabla j}_{t}\textbf{s}_{t}(\textbf{r})\cdot[\boldsymbol{\nabla}\times\textbf{j}_{t}](\textbf{r})$, where $t=0(1)$ refers to the isoscalar (isovector) character of the coupling constants, $\textbf{s}_{t}(\textbf{r})$ is the spin density, $\textbf{T}_{t}(\textbf{r})$ represents the vector kinetic density, and $\textbf{j}_{t}(\textbf{r})$ represents the current density. As in our previous work~\cite{(Sas22c),(Bon23c)}, we parameterised coupling constants $C^{s}_{t}$ and $C^{T}_{t}$ by the standard isoscalar and isovector Landau parameters, $g_{0}$ and and $g'_{0}$~\cite{(Ben02d)}, and we used the values of $g_{0} = 0.4$ and $g'_{0} = 1.7$.  For the functional UNEDF1, these values of the Landau parameters correspond to the Skyrme-functional coupling constants of $C^{s}_{0} = 30.555$ and $C^{s}_{1} = 129.858$\,MeV\,fm$^3$. In addition, the coupling constants $C^{T}_{t}$, $C^{j}_{t}$, and $C^{\nabla j}_{t}$ were kept fixed at their values given by the gauge symmetry conditions~\cite{(Dob95b),(Per04c)} and  $C^{\Delta s}_{t}$ were set to zero to avoid known time-odd instabilities~\cite{(Hel12)}.

As in our previous work~\cite{(Bon23c)}, in the pairing channel, we used the mixed pairing force adjusted in~\cite{(Kor12b)}, and we increased the strength parameters by 20\% to account for replacing the original Lipkin-Nogami (LN) method with the HFB approach. Within the LN method and the BCS model, see, e.g.,~\cite{(Bender00)}, the approximate particle-number projected spectral gaps are given by $\langle v^{2}\Delta\rangle^{(\text{LN})}_{q}=\langle v^{2}\Delta\rangle_{q}+\lambda_{2,q}$ and $\langle uv\Delta\rangle^{(\text{LN})}_{q}=\langle uv\Delta\rangle_{q}+\lambda_{2,q}$, where $q\in\{p,n\}$, and  are larger by the LN second-order coefficient $\lambda_{2,q}$  than the corresponding BCS gaps. Although the discussion in~\cite{(Bender00)} centres on pairing gaps within the BCS approximation, a similar conclusion holds for the HFB approach used here because, within a canonical basis~\cite{(Rin80)}, the BCS approximation is equivalent to the HFB approach. Since the calculated pairing gaps increase with the pairing strengths, the LN contributions of $\lambda_{2}$ to the spectral gaps can be compensated by increasing the HFB pairing strengths. A 20\% increase was chosen so that our calculated HFB values of the odd-even mass staggering (section \ref{pairing}) are, on average, compatible with experimental data. 

We performed calculations using the code {\sc hfodd} (v3.16p)~\cite{Dobaczewski_2021,(Dob25a)} while conserving parity symmetry. The numerical values of the results determined in this study are available in the Supplemental Material~\cite{supp-SnGd}. The 3D Cartesian HO basis, characterised by equal HO frequencies along the three Cartesian directions, $\omega_x=\omega_y=\omega_z\equiv\omega_0$, was employed. The spherical single-particle basis was established by limiting the number of HO quanta in the three Cartesian directions, $n_x$, $n_y$, and $n_z$, so that $n_x+n_y+n_z\leq{}N_0=16$, where $N_0$ represents the maximum number of spherical HO shells included. The sphericity of the HO basis did not impede arbitrary deformations of self-consistent solutions. However, in this application, the 3D code was run in a mode that enforced the axial symmetry of the solutions.
  
As it turned out, almost all calculations performed for both tags (prolate and oblate) and for all 160 odd-$N$ and 147 odd-$Z$ nuclei (figure~\ref{fig1}) converged well. Isotopes for which the oblate-tag calculations did not converge were $^{115}$Ba and $^{157}$Sm. Similarly, the prolate-tag calculations for $^{115}$I, $^{145}$I, $^{127}$Cs and $^{157}$Pm did not converge either. 

\begin{figure}[ht]
\begin{center}
\includegraphics[width=0.9\textwidth]{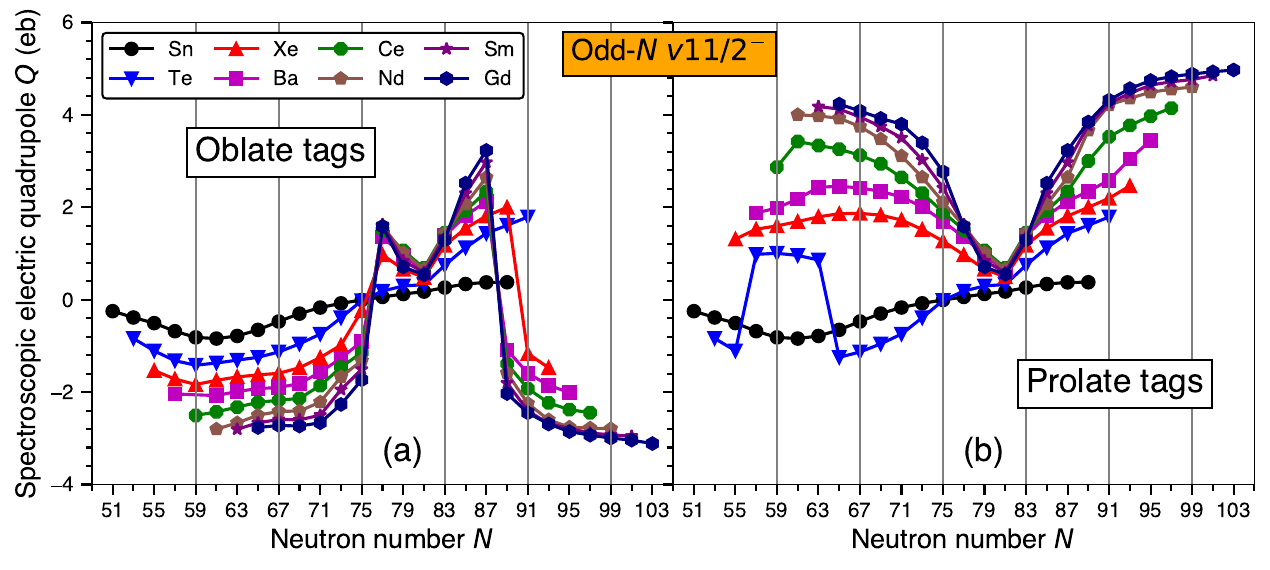}
\end{center}
\caption{\label{fig3} Spectroscopic electric quadrupole moments in the $\nu11/2^{-}$ configurations of odd-$N$ nuclei as functions of the neutron number $N$, obtained from the oblate (a) and prolate (b) tags.
}
\end{figure}
\begin{figure}[ht]
\begin{center}
\includegraphics[width=0.9\textwidth]{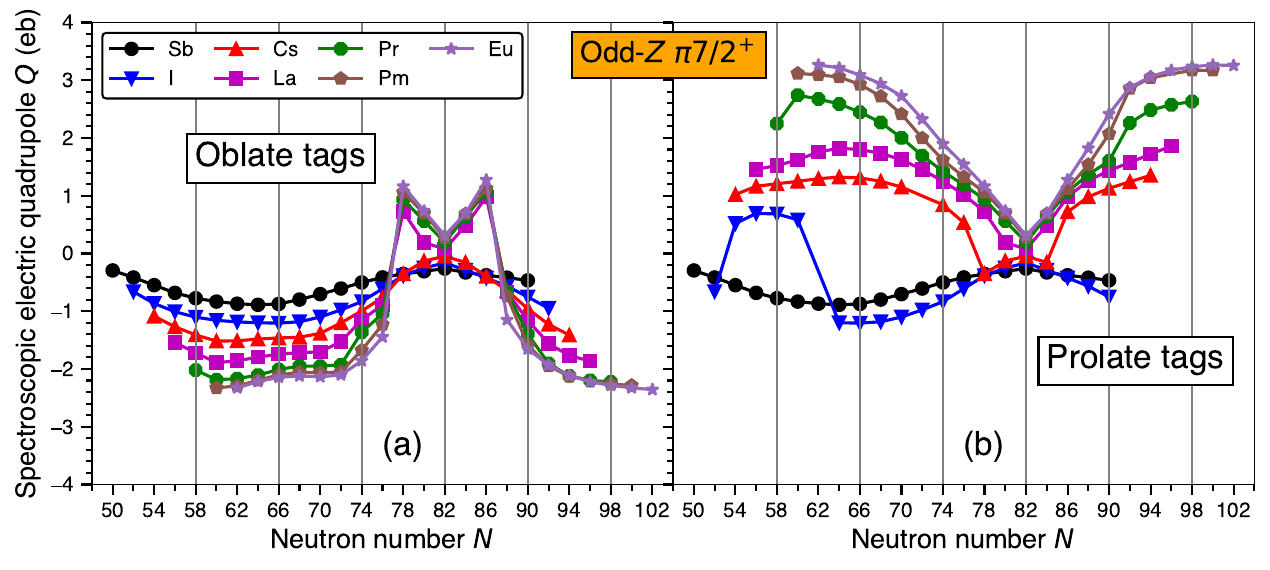}
\end{center}
\caption{\label{fig3a} Same as in figure~\ref{fig3} but for the $\pi7/2^{+}$ configurations of odd-$Z$ nuclei. 
}
\end{figure}
 Panels of figure~\ref{fig3} (figure~\ref{fig3a}) show the neutron number dependence of the spectroscopic electric quadrupole moments of odd-$N$ (odd-$Z$) nuclei obtained from the oblate (a) and prolate (b) tags. Missing points indicate the non-converged calculations. Before discussing the deformation characteristics of the studied nuclei, we draw the reader's attention to the fact that apart from a few points where the obtained fixed-tag results jump between prolate and oblate shapes, they vary remarkably smoothly. This feature of the tagged calculations is helpful because, away from the jumps, it allows us to replace all missing non-converged results with the corresponding interpolations between the converged neighbours. Interpolated results are indicated in the Tables of the Supplemental Material~\cite{supp-SnGd} and included in all figures below.

\begin{figure}[ht]
\begin{center}
\includegraphics[width=0.9\textwidth]{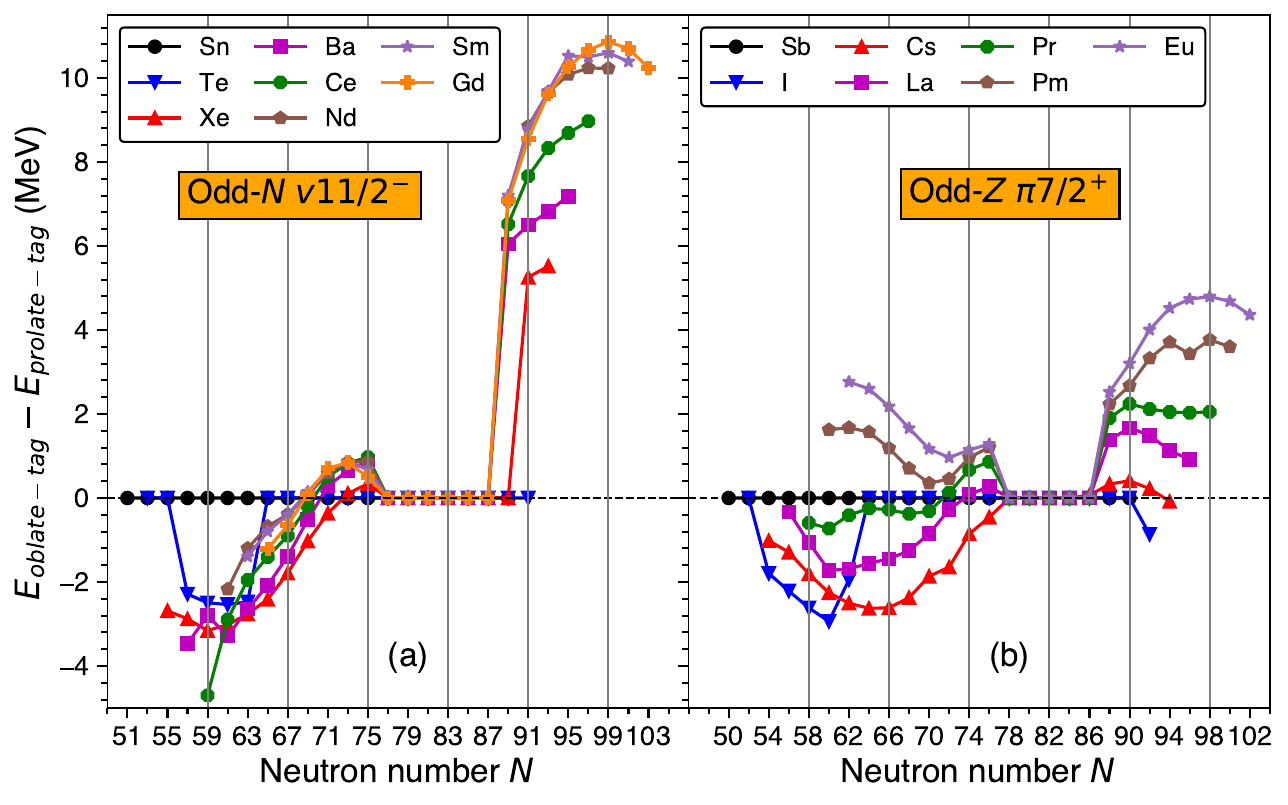}
\end{center}
\caption{\label{fig2-a-b} Energy differences between oblate-tag and prolate-tag minima for odd-$N$ (a) and odd-$Z$ (b) nuclei. The energy difference $E_{\text{oblate-tag}}-E_{\text{prolate-tag}}$ for $^{145}$I isotope ($N=92$) was linearly interpolated using the $E_{\text{prolate-tag}}$ of $^{143}$I and $^{147}$I with $N=90$ and $N=94$, respectively, the latter being outside the set of isotopes defined in figure~\protect\ref{fig1}.
}
\end{figure}
In figure~\ref{fig2-a-b}, for odd-$N$ (a) and odd-$Z$ (b) solutions, we show the energy differences between the oblate-tag and prolate-tag results. Vanishing values of those differences indicate cases where the prolate and oblate tags led to the unique (prolate or oblate) global minima. Results of that kind were obtained for (i) all Sn isotopes, (ii) all Te isotopes apart from   $57\leq{N}\leq63$, and all odd-$N$ elements with $77\leq{N}\leq87$. Similarly, they were also obtained for (i) all Sb isotopes, (ii) all I isotopes apart from $50\leq{N}\leq62$ and $N=92$, and (iii) all odd-$Z$ elements with $78\leq{N}\leq 86$. Other odd nuclei exhibit two different minima with positive (negative) oblate-prolate energy differences, which implies prolate (oblate) global minima. 

\begin{figure}[ht]
\begin{center}
\includegraphics[width=0.95\textwidth]{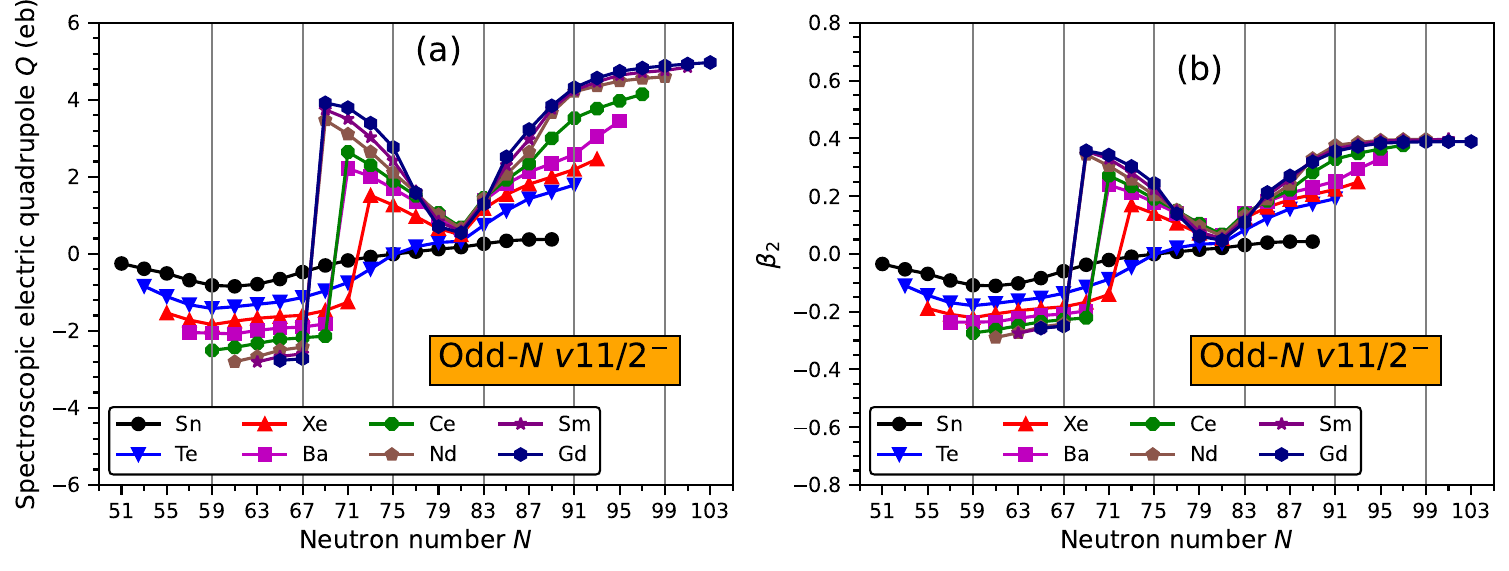}
\end{center}
\caption{\label{fig4} Same as in figure~\protect\ref{fig3} but for the global minima (a) and the corresponding Bohr parameters $\beta_{2}$ (b) calculated using Eq.~\eqref{Bohr_params}.
}
\end{figure}
\begin{figure}[ht]
\begin{center}
\includegraphics[width=0.95\textwidth]{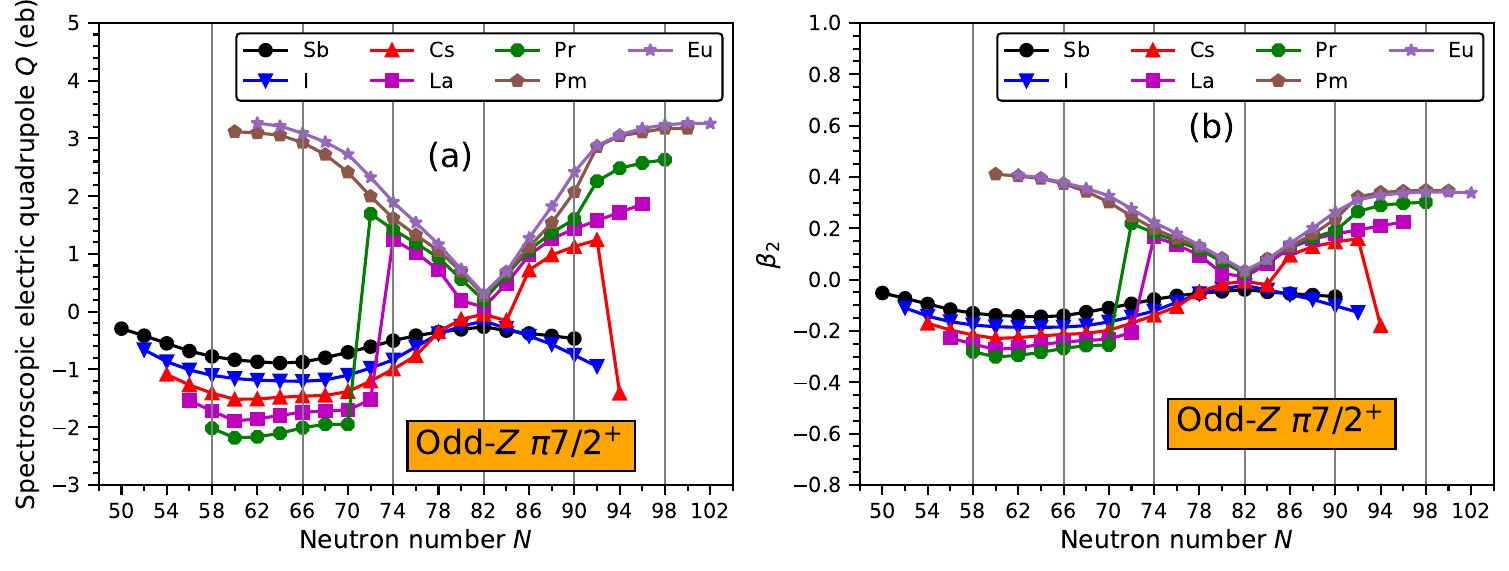}
\end{center}
\caption{\label{fig5} Same as in figure~\protect\ref{fig3a} but for the global minima (a) and the corresponding Bohr parameters $\beta_{2}$ (b) calculated using Eq.~\eqref{Bohr_params}.
}
\end{figure}
It should be noted that this analysis is limited to axial shapes. Consequently, for small oblate/prolate energy differences or small deformations, triaxial shapes and the generator coordinate mixing thereof leading to shape coexistence may be relevant, which is beyond the scope of the current study. In the following, we present results solely for the global minima (unique, prolate-tag, or oblate-tag), and the Tables in the Supplemental Material~\cite{supp-SnGd} provide information on the type of minimum selected.

In figures~\ref{fig4}(a) and~\ref{fig5}(a), we present the spectroscopic quadrupole moments $Q$ calculated at the global minima of odd-$N$ and odd-$Z$ nuclei, respectively. For illustration, in figures~\ref{fig4}(b) and~\ref{fig5}(b), we also present the corresponding Bohr $\beta_2$ parameters, calculated from the values of $Q$ in an analogous way as they are routinely determined from the measured spectroscopic quadrupole moments, that is,
\begin{equation}
\label{Bohr_params}
    \beta_{2} = \frac{\sqrt{5\pi}}{3ZR^{2}_{0}}\frac{Q}{\left(C^{II}_{II,20}\right)^2},
\end{equation}
where $R_{0}$ is the standard parameterisation of the nuclear geometric radius in terms of the mass number $A$, $R_{0}=1.2A^{1/3}$\,fm, and $C^{II}_{II,20}$ are the Clebsch-Gordan coefficients, in our case reading $C^{{11}/{2}\,{11}/{2}}_{{11}/{2}\,{11}/{2},20}=\sqrt{{55}/{91}}$ and $C^{{7}/{2}\,{7}/{2}}_{{7}/{2}\,{7}/{2},20}=\sqrt{{7}/{15}}$.

For the Sn and Te isotopic chains, figure~\ref{fig4} shows a smooth transition between oblate and prolate shapes with the neutron numbers varying from the magic number 50 to beyond the magic number 82. Conversely, at $N=67-69$, the isotopic chains of heavier elements exhibit abrupt transitions between oblate and prolate shapes. As explained in~\cite{(Bon23c)}, abrupt transitions occur because configurations that start as holes (particles) at the lower end of an open shell, at the higher end must become particles (holes). Figure~\ref{fig4} also illustrates that the closer (farther) the proton number is to the magic number 50, the abrupt transition occurs at larger (smaller) neutron numbers. This generic behaviour arises from the interaction between the shape polarisation of protons and neutrons. Figure~\ref{fig4} also shows that the same signs of the spectroscopic quadrupole moments are obtained when we follow the $\nu11/2^{-}$ configurations across the neutron $N=82$ magic number, as on both sides, this configuration has a hole character. Analogous conclusions can be drawn for the $\pi7/2^{+}$ configurations in odd-$Z$ nuclei shown in figure~\ref{fig5}.

\begin{figure}[ht]
\begin{center}
\includegraphics[width=0.7\textwidth]{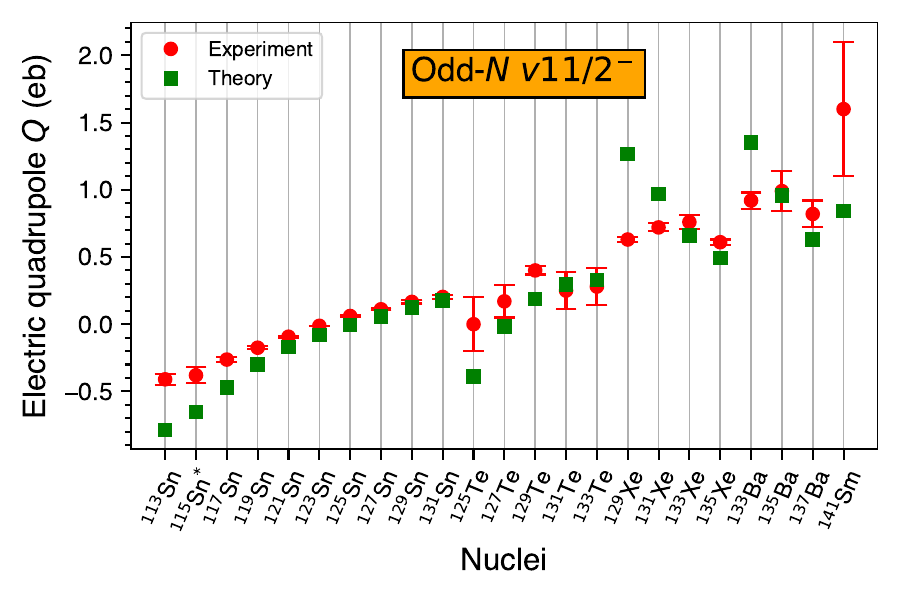}
\end{center}
\caption{\label{fig6} Comparison between the calculated electric quadrupole $Q$ moments of the $\nu11/2^{-}$ configurations in odd-$N$ nuclei and the available experimental data~\cite{Stone2014, Stone2016, Stone2021, Yordanov2020}. The asterisk ($^{*}$) represents the case where the sign of the electric quadrupole moment was not determined experimentally, and thus the calculated sign was adopted.}
\end{figure}
\begin{figure}[ht]
\begin{center}
\includegraphics[width=0.7\textwidth]{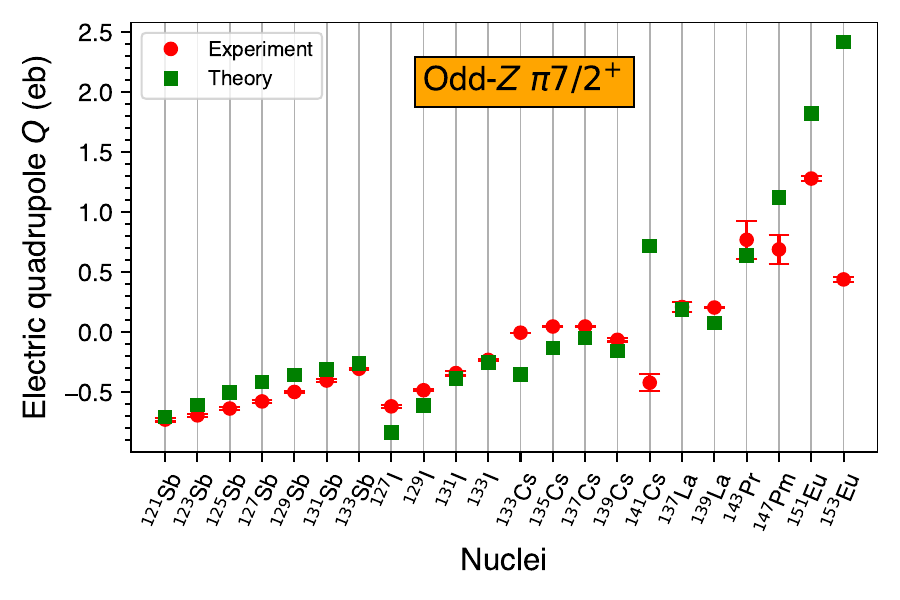}
\end{center}
\caption{\label{fig7} Same as in figure~\protect\ref{fig6} but for the $\pi7/2^{+}$ configurations in odd-$Z$ nuclei~\cite{Stone2014, Stone2016, Stone2021, Alkhazov1992, Haiduke2006,(Lec21a), (Lec23)}.
}
\end{figure}
\begin{figure}[ht]
\begin{center}
\includegraphics[width=0.7\textwidth]{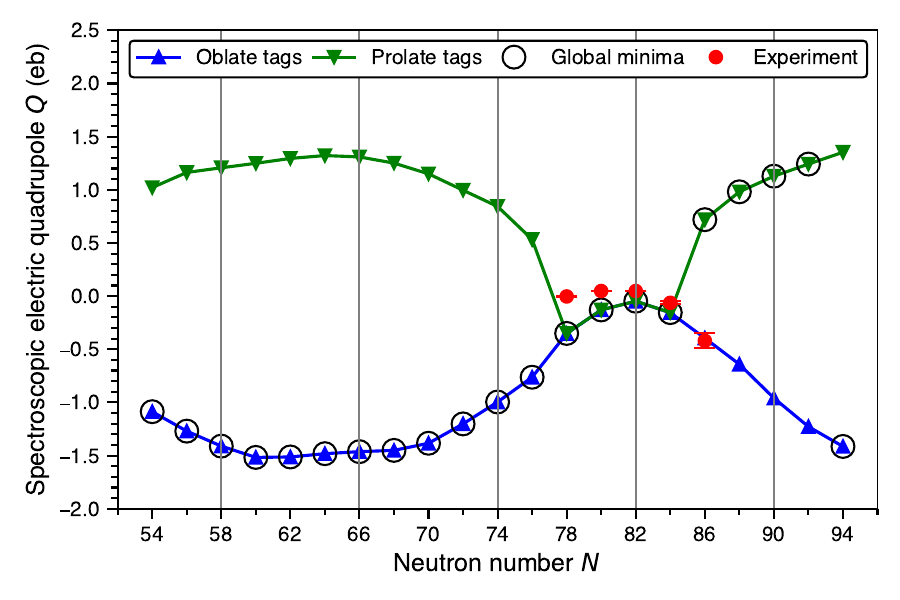}
\end{center}
\caption{\label{fig8} Evolution of the quadrupole moments $Q$ with neutron numbers $N$ obtained in Cs isotopes for oblate (up triangles) and prolate (down triangles) tags. Open and full circles represent the global minima and experimental data, respectively.
}
\end{figure}

Figure~\ref{fig6} (figure~\ref{fig7}) shows that for odd-$N$ (odd-$Z$) nuclei, the calculated spectroscopic quadrupole moments agree reasonably well with experimental data~\cite{Gray}. It is worth noting that the results do not depend on the time-odd mean fields and, thus, on the specific values of the Landau parameters used here. In the case of $^{141}\text{Cs}$, the sign of Q obtained in the calculations is opposite to that determined experimentally. As shown in figure~\ref{fig8}, for the neutron numbers between 78 and 84, the results obtained for global minima agree on the trend of quadrupole moments. In addition, the spectroscopic quadrupole moment of $^{141}\text{Cs}$ ($N = 86$) obtained from the oblate tag matches the experimental value spectacularly well. This shows that in the Cs isotopes, theoretical calculations predict the abrupt transition from oblate to prolate shapes two neutron numbers too early. However, we should keep in mind that in $^{141}\text{Cs}$, the oblate-prolate energy difference, figure~\ref{fig2-a-b}, is as small as 64\,keV. We also note that another large deviation between theory and experiment, obtained in $^{153}\text{Eu}$, cannot be explained.

\begin{figure}[ht]
\begin{center}
\includegraphics[width=1.0\textwidth]{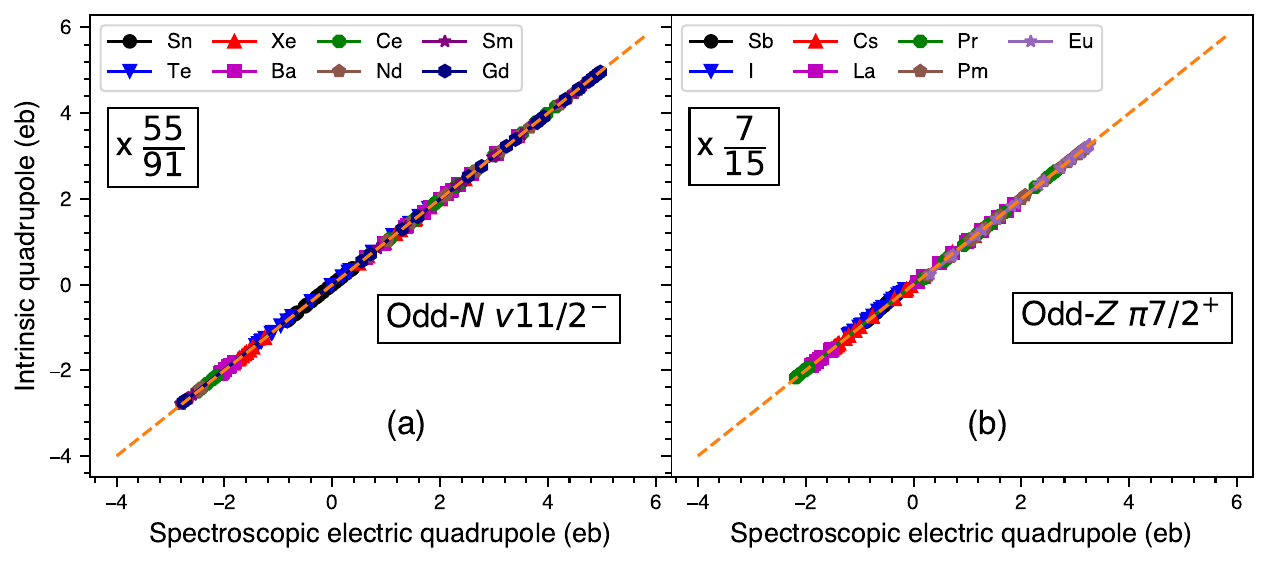}
\end{center}
\caption{\label{fig9} Calculated spectroscopic quadrupole moments $Q$ compared with estimates~(\protect\ref{large}) for the $\nu11/2^{-}$ configurations in odd-$N$ nuclei (a) and $\pi7/2^{+}$ configurations in odd-$Z$ (b) nuclei.}
\end{figure}
We investigated the importance of the AMP in determining the spectroscopic electric quadrupole moments by testing the validity of the large-axial-deformation (or rigid-rotor) approximation\footnote{See~\protect\cite{(Rin80)}, Eq.~(11.143).}, already used in~\eqref{Bohr_params}. It allows for estimating the spectroscopic quadrupole moments $Q^{\text{spec}}_{\text{est}}$ from the calculated intrinsic quadrupole moments $Q^{\text{intr}}_{20}$ as,
\begin{equation}
\label{large}
    Q^{\text{spec}}_{\text{est}}=\left(C^{II}_{II,20}\right)^{2}\times Q^{\text{intr}}_{20}.
\end{equation}
\begin{figure}[ht]
\begin{center}
\includegraphics[width=1.0\textwidth]{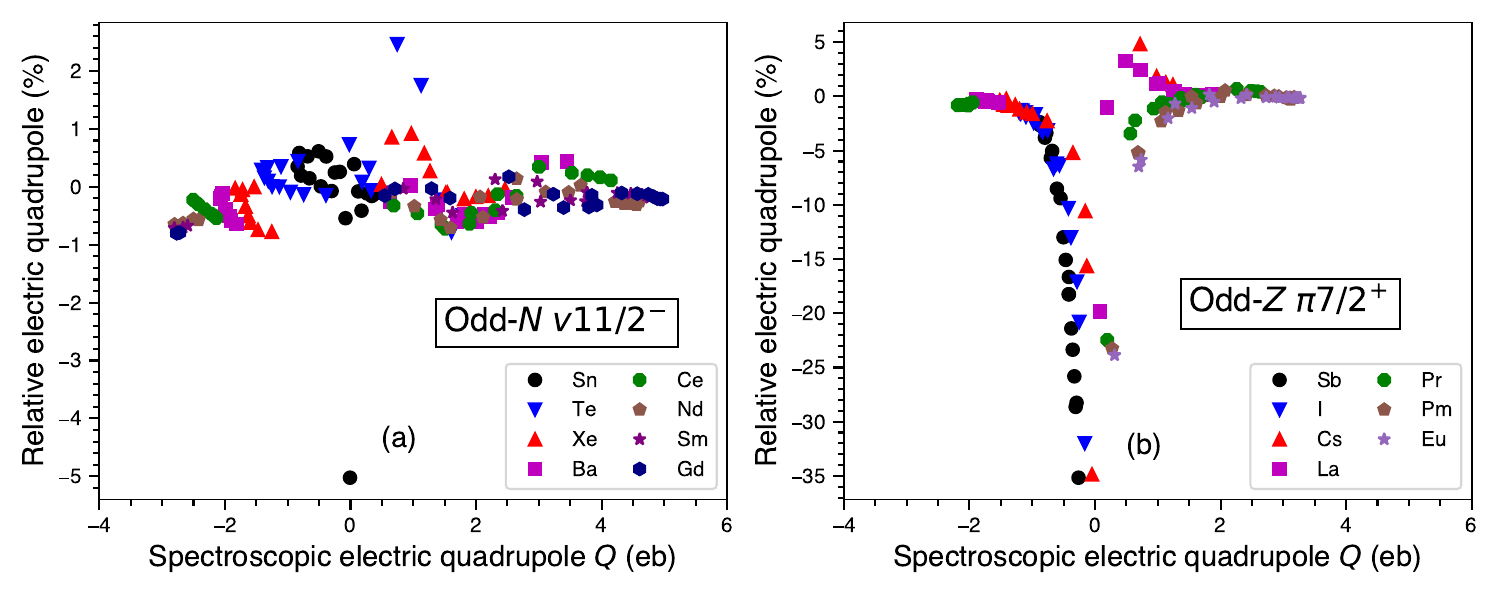}
\end{center}
\caption{\label{fig11} Same as in figure~\protect\ref{fig9} but for the relative electric quadrupole moments, see text.}
\end{figure}
In panels (a) and (b) of figure~\ref{fig9}, for the $\nu11/2^{-}$ and $\pi7/2^{+}$ configurations in odd-$N$ and odd-$Z$ nuclei, respectively, we compare the calculated spectroscopic electric quadrupole moments $Q$ with their rigid-rotor approximations~(\ref{large}). We observe that, when plotted in the large scale used in the figure, the approximations seem to be perfect. However, to visualise their precision at small deformations, in figure~\ref{fig11}, we show the corresponding relative percentage deviations, $\left(Q^{\text{spec}}_{\text{est}}-Q\right)/Q$. Here we see that for the $\nu11/2^{-}$ configurations in odd-$N$ nuclei, at all deformations, the rigid-rotor approximation is precise up to about 1\%, whereas for the $\pi7/2^{+}$ configurations in odd-$Z$ nuclei it deteriorates at small deformations to about 35\%. Figure~\ref{fig11} may serve as a guide for employing the rigid-rotor approximation rather than conducting the exact AMP in weakly deformed nuclei. 

\begin{figure}[ht]
\begin{center}
\includegraphics[width=0.8\textwidth]{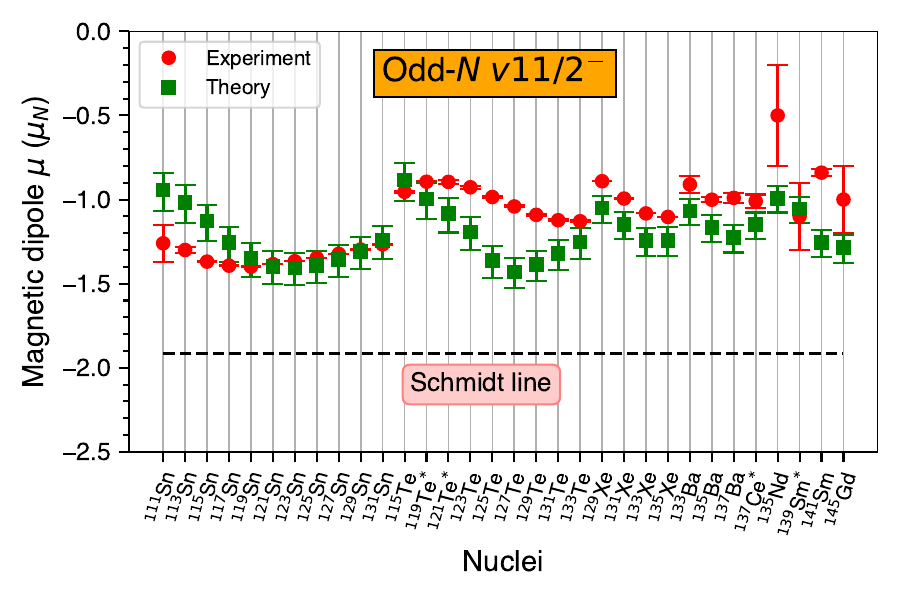}
\end{center}
\caption{\label{fig12} Comparison between the calculated spectroscopic magnetic dipole moments $\mu$ of the $\nu11/2^{-}$ configurations in odd-$N$ nuclei with experimental data~\cite{Stone2014, Stone2019, Yordanov2020, Stone2020}. The uncertainties of the theoretical values were obtained by recalculating the theoretical results for the Landau parameters $g'_{0}=1.3$ and $2.1$, which correspond to the uncertainty of $g'_{0}=1.7(4)$ estimated in~\protect\cite{(Sas22c)}. The asterisks ($^{*}$) represent the cases where the signs of the magnetic dipole moments were not determined experimentally, and the calculated signs were adopted.
}
\end{figure}
\begin{figure}[ht]
\begin{center}
\includegraphics[width=0.8\textwidth]{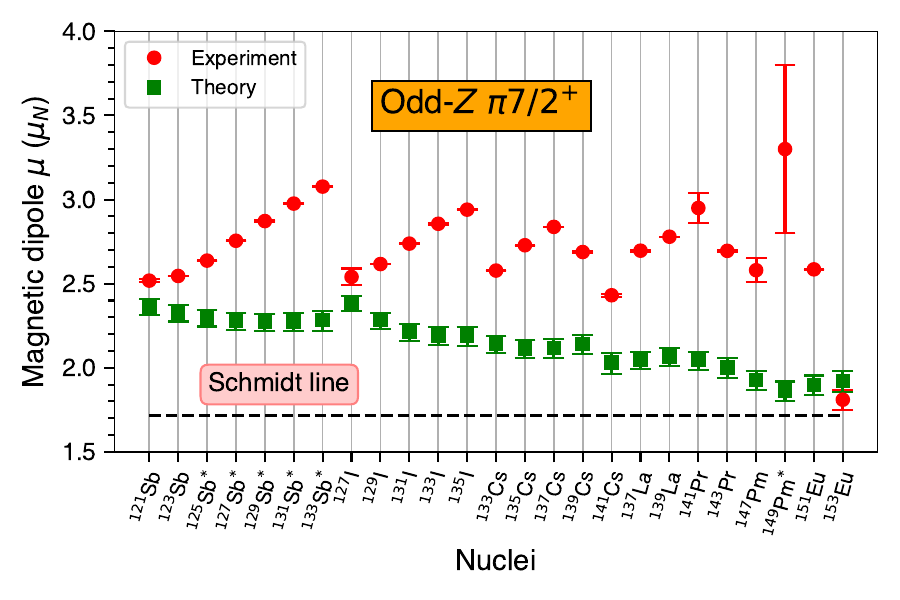}
\end{center}
\caption{\label{fig13} Same as in figure~\protect\ref{fig12} but for the $\pi7/2^{+}$ configurations in odd-$Z$ nuclei
~\cite{(Lec23)}.}
\end{figure}
In figures~\ref{fig12} and~\ref{fig13}, which correspond to the $\nu11/2^{-}$ and $\pi7/2^{+}$ configurations in odd-$N$ and odd-$Z$ nuclei, respectively, we compare the calculated spectroscopic magnetic dipole moments $\mu$ with experimental data. For the spin-aligned configuration, $\nu11/2^{-}$, the agreement with the data is reasonably good~\cite{Gray}; however, the experimental trend is not reproduced in the Te isotopes. In contrast, the values and trends observed for the spin-anti-aligned configuration, $\pi7/2^{+}$, do not align well with the data. It should be noted that in~\cite{(Lec23)}, the Sb experimental results were reproduced by adjusting the values of three effective $g$-factors in both shell-model and {\it ab initio} calculations. However, the physical rationale behind using these specific values remains unclear.

\begin{figure}[ht]
\begin{center}
\includegraphics[width=1.0\textwidth]{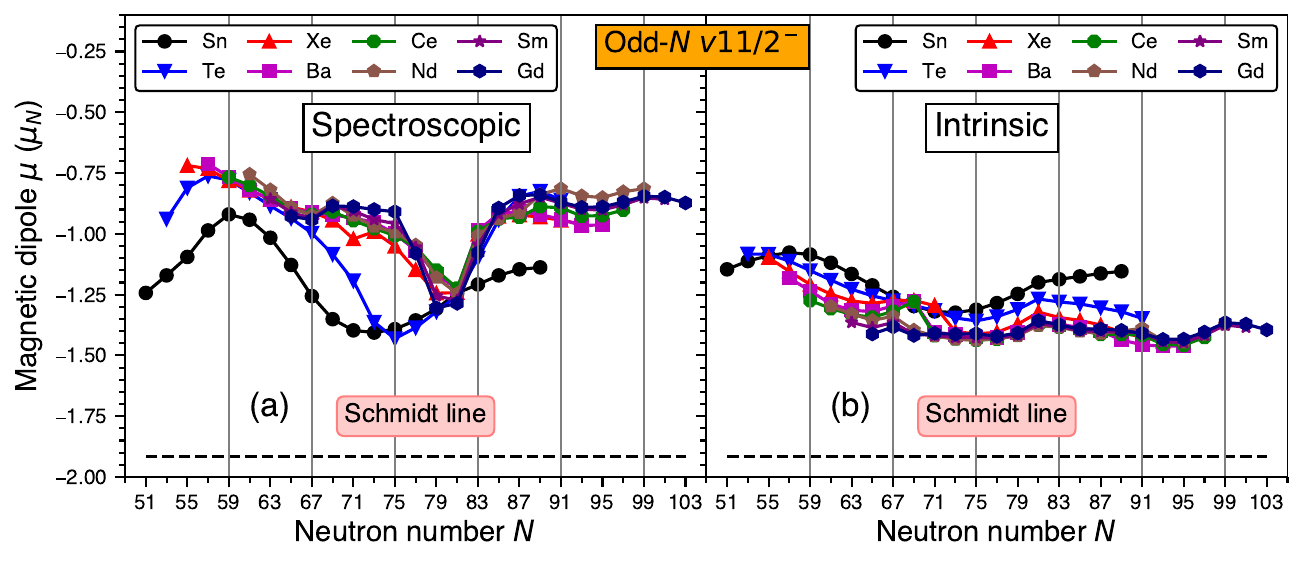}
\end{center}
\caption{\label{fig14} Calculated intrinsic magnetic dipole moments compared with the spectroscopic ones for the $\nu11/2^{-}$ configurations in odd-$N$ nuclei.
}
\end{figure}

\begin{figure}[ht]
\begin{center}
\includegraphics[width=1.0\textwidth]{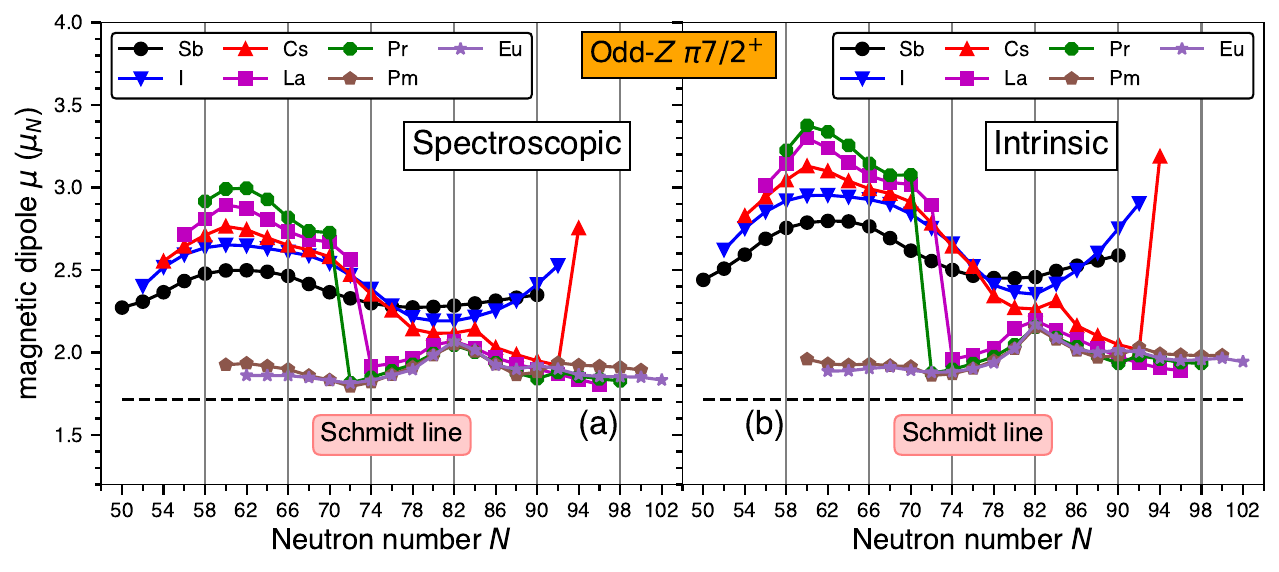}
\end{center}
\caption{\label{fig15} Same as in figure~\protect\ref{fig14} but for the $\pi7/2^{+}$ configurations in odd-$Z$ nuclei.
}
\end{figure}

As shown in figures~\ref{fig14} and~\ref{fig15}, which compare the spectroscopic and intrinsic values of the calculated magnetic dipole moments, the restoration of rotational symmetry (AMP) is notable, with trends—particularly for $\nu11/2$—and values altered accordingly. 

\section{Precision of calculations}\label{Accuracy}

\subsection{Harmonic oscillator basis}\label{HO}

The calculations presented in this work were obtained using a spherical harmonic oscillator (HO) basis up to $N_0=16$ shells, where the value of the "physical" frequency  $\hbar \omega_0= f \times$ 41 MeV$\cdot$ A$^{1/3}$ was established by setting $f=1.2$~\cite{(Dob97c)}. To examine how the spectroscopic moments differ when altering the properties of the spherical HO basis, we selected $^{167}$Gd, one of the most deformed nuclei in the region studied in this work. Therefore, it is reasonable to anticipate a similar range of deviations for the other nuclei investigated. 

\begin{figure}[ht]
\begin{center}
\includegraphics[width=0.70\textwidth]{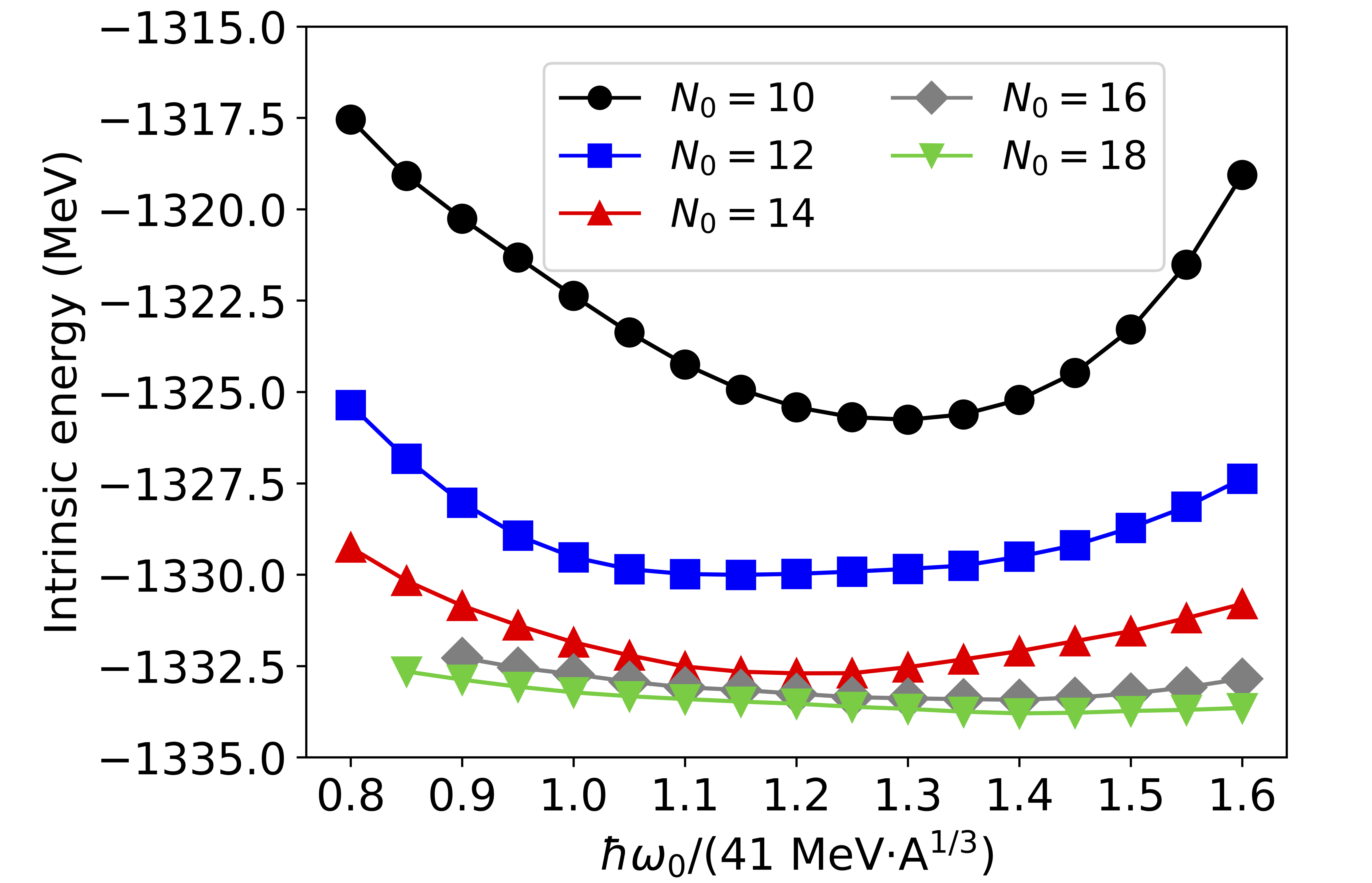}
\end{center}
\caption{\label{fig_HO_ener}Ground state energies of the oblate-tag minima in the ${11/2}^-$ state of $^{167}$Gd as functions of the HO basis frequency $\hbar\omega_0$.
}
\end{figure}

The choice of the number of shells to include in the calculations depends on the convergence of results. Once a sufficient number of basis states are considered, increasing the basis size only slightly impacts the system's properties. However, with a larger basis comes a more significant computational burden, making it essential to find the optimal set of states to balance precision and workload. To illustrate this analysis, in Figure~\ref{fig_HO_ener}, we plotted the energy of the oblate-tag minimum of the ${11/2}^-$ state for the $^{167}$Gd nucleus as a function of different numbers of shells and HO frequencies. We can see that using a basis larger than $N_0=14$ has a minor effect on reducing the oblate minimum energy. For instance, in this example, the variation between $N_0=16$ and $N_0=18$ is 272\,keV. Even more interestingly, at larger values of $N_0$, the results are nearly insensitive to the choice of $\hbar\omega_0$, cf.~\cite{(Dob97c)}.

\begin{figure}[ht]
\begin{center}
\includegraphics[width=0.90\textwidth]{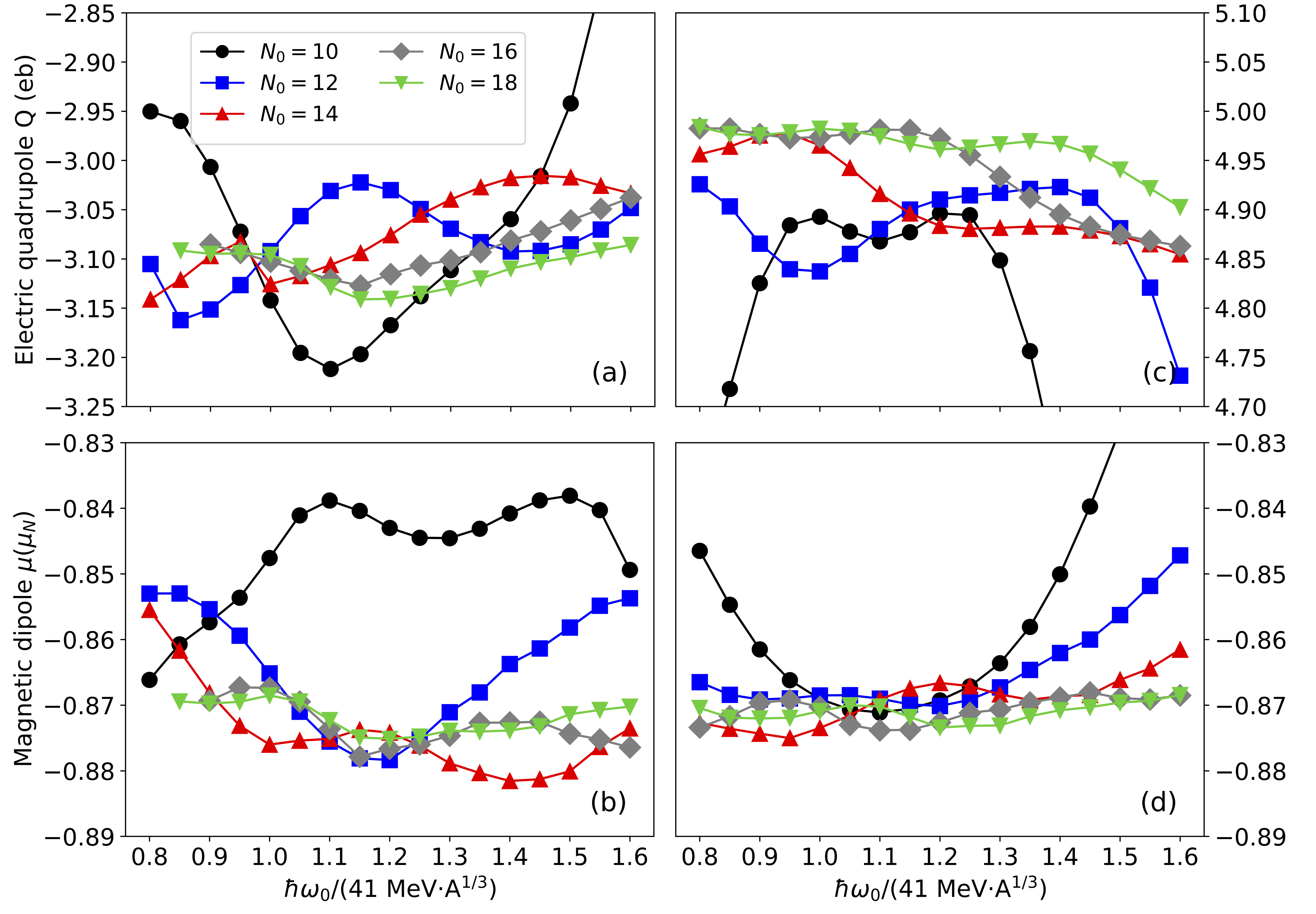}
\end{center}
\caption{\label{fig_HO_test}(colour online) Spectroscopic electric quadrupole moments $Q$ (upper panels) and magnetic dipole moments $\mu$ (lower panels) for the oblate (left panels) and prolate (right panels) minima of $^{167}$Gd using a spherical HO basis with different numbers of shells, $N_0$, and HO frequencies, $\omega_0$.
}
\end{figure}

Figure~\ref {fig_HO_test} presents the electric quadrupole moments $Q$ (top panels) and magnetic dipole moments $\mu$ (bottom panels) determined at the oblate-tag minimum (left panels) and prolate-tag minimum (right panels) for different values of $N_0$ and $\hbar\omega_0$. We notice that the results obtained at $N_0 \leq 14$ fluctuate significantly before stabilising. Specifically, at $N_0=16$, the standard deviations are in the range of $\Delta{Q}\simeq0.03-0.05$\,b and  $\Delta{\mu}\simeq0.002-0.003$\,$\mu_N$. To conclude, if we consider those deviations as typical uncertainties related to the basis truncation, the related error bars in all figures presented in this study would be invisible. For instance, the size of the symbols representing values of $Q$ ($\mu$) in figure~\ref{fig6} (figure~\ref{fig12}) is 0.1\,b (0.1\,$\mu_N$), which is significantly larger than the numerical uncertainty estimated here.

\subsection{Pairing strength}\label{pairing}

\begin{figure}[ht]
\begin{center}
\includegraphics[width=0.89\textwidth]{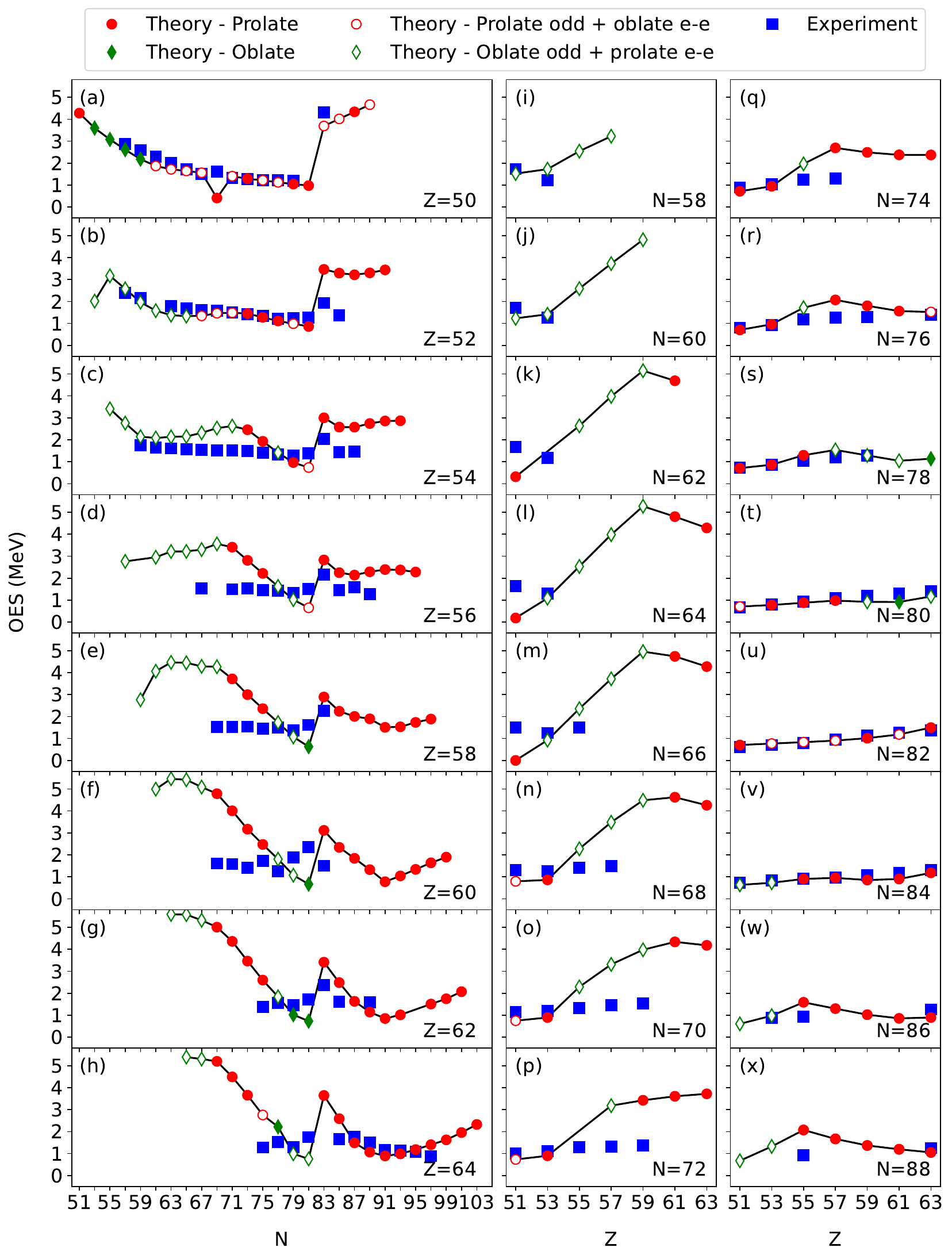}
\end{center}
\caption{\label{fig_oes} Calculated neutron (panels a-h) and proton (panels i-x) odd-even mass staggering (OES) corresponding to the energies of the blocked $\nu11/2^{-}$ and $\pi7/2^{+}$ configurations, respectively, compared with the available experimental data~\cite{ENSDF.230918}. }
\end{figure}
Throughout this work, we observed a reasonable agreement with experimental data for the electric quadrupole moments $Q$. However, the calculated spectroscopic magnetic dipole moments $\mu$ of the ${\pi7/2^{+}}$ configuration poorly reproduce the data. Since the magnetic dipole moments strongly depend on both the single-particle properties and the polarisation of the paired core, in this section, we critically evaluate the adjustment of the pairing strengths in the studied nuclei.

To assess the quality of the proton and neutron pairing strengths, we determined the proton and neutron odd-even mass staggering (OES), the difference in binding energy between a given odd nucleus and the average of its even-even neighbours. To do this, we calculated the ground-state energies of all even-even neighbours using the same pairing strengths as those employed in odd nuclei. It is worth noting that the OES values obtained in this manner are not identical to those typically calculated to evaluate the ground-state pairing energies. Indeed, by utilising the binding energies of specific configurations, regardless of their positions in the spectra of odd nuclei, we assess the combined effects of the pairing strengths and isomer excitation energies. In contrast, the standard ground-state values mix the effects of the pairing strengths and configuration changes. 

We analysed the pairing strengths using the global or unique minima in odd nuclei, as detailed in section~\ref{Results}, which may be either prolate or oblate. To enhance the reliability of our analysis, we consistently selected (i) identical shapes for the two even-even neighbours and (ii) either the same or opposite shapes for the odd nuclei and their even-even neighbours. The experimental data were obtained by subtracting the isomer excitation energies of the odd nuclei~\cite{ENSDF.230918} from their corresponding ground-state binding energies~\cite{(Wan21)}. Our results are depicted in figure~\ref{fig_oes}, where panels a-h (i-x) correspond to the neutron (proton) OES. The symbols in the legend represent the shapes of the odd nuclei as related to those of the even-even neighbours.

Our results for the neutron OES show good agreement with data, particularly for the Sn isotopes. We replicate the jumps in the OES around ${N=82}$, associated with the shell closure, although, at $Z>50$, the jumps in the experimental values are less pronounced. As we examine heavier isotopic chains, our calculations do not fully align with the behaviour observed in the experiment. However, concerning the overall magnitude of the OES, the disagreement between theoretical and experimental values does not seem systematic. For the proton OES, our results demonstrate better agreement with the data.

Adjusting the pairing strengths would imply shifting the theoretical lines toward higher or lower values, which would not address the inconsistencies with the OES experimental values. Therefore, we conclude that the pairing strengths utilised in this work are reasonably calibrated, and there is no compelling reason to alter them. Although the OES analysis does not clarify the inconsistencies in our results for the magnetic dipole moments $\mu$, it eliminates the pairing strengths as a potential factor. Additionally, this maintains the generality of DFT calculations without necessitating parameter adjustments for specific regions of the nuclear chart, which is one of the commendable features of this work.

\subsection{Particle-number projection}\label{PNP}

\begin{figure}
\centering
\includegraphics[width=0.5\textwidth]{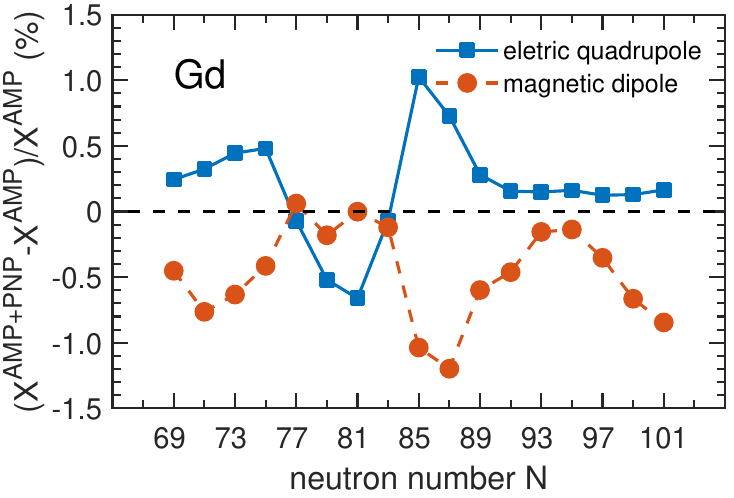}
\caption{Relative differences (in \%) between the electric quadrupole ($X\equiv{Q}$) and magnetic dipole ($X\equiv\mu$) moments calculated with and without the PNP. The plot shows the results obtained for the $\nu11/2^{-}$ configurations in odd-N isotopes of gadolinium.
\label{Gd_PNP}
}
\end{figure}
In figure~\ref{Gd_PNP}, we display the percentage deviations between the spectroscopic (AMP) magnetic dipole moments $\mu$ and electric quadrupole moments $Q$ calculated with and without PNP. The PNP results were obtained using 9 Gauss-Tchebyshev integration nodes~\cite{Sheikh_2021, Dobaczewski_2021} for both proton and neutron gauge angles. Tests were conducted for the $\nu11/2^{-}$ configurations in odd-$N$ isotopes of gadolinium, yielding results that differ by only about 1\% from those obtained without the PNP. This finding supports the sole use of AMP when determining electromagnetic moments. Given that the PNP increases the computational time by around two orders of magnitude (9$\times$9), this fact greatly enhances the feasibility of this work's large-scale nuclear DFT calculations.

\section{Conclusions and outlook}\label{Conclusions|}

In this work, we presented the nuclear DFT description of electric quadrupole and magnetic dipole moments in odd open-shell nuclei, spanning the elements from tin to gadolinium. By aligning the angular momenta along the axial symmetry axis, incorporating the signature and time-reversal breaking as well as time-odd mean fields, and restoring rotational symmetry, we have proven these to be critical elements of the description.

By tagging the quasiparticle states with the single-particle wave functions determined in $^{100}$Sn, we were able to trace two specific paired configurations of $\nu11/2^{-}$ in odd-$N$ and $\pi7/2^{+}$ in odd-$Z$ nuclei. We achieved reasonable agreement for both configurations between the calculated electric quadrupole moments and experimental data. Furthermore, for the parallel coupling of spin and orbital angular momenta in the $\nu11/2^{-}$ configurations, the calculated magnetic dipole moments also align well with data, particularly in the Sn and Xe isotopes. In contrast, neither the values nor the trends in the experimental data could be reproduced for the antiparallel coupling in the $\pi7/2^{+}$ configurations. We also examined the precision of the results against the truncation of the HO basis, strengths of the pairing interactions, and restoration of particle-number symmetry, concluding that none of these factors had a meaningful impact on the results. In comparison to restoring the particle-number symmetry and angular momentum symmetry for triaxial shapes, the methodology employed here reduced computational costs by four orders of magnitude and enabled efficient large-scale calculations.

A substantial disagreement exists between the calculated and experimental magnetic dipole moments of the $\nu7/2^{+}$ configurations, necessitating further investigation. On the one hand, attributing this disagreement to adjusted effective $g$-factors does not further scientific understanding. On the other hand, potential physical explanations for the disagreement identified here must be weighed against the agreement observed for the $\pi11/2^{-}$ configurations. At this stage, we can argue that three promising avenues for further research exist. Firstly, one could aim to incorporate more sophisticated time-odd mean fields beyond those provided by the simple isovector spin-spin terms employed here. Secondly, in the region of the studied nuclei, the effects of triaxiality and collectivity may be significant. Thirdly, as suggested long ago (see, e.g.,~\cite{(Pad65),(Che69)}) and recently demonstrated in~\cite{(Miy24)}, the one-body magnetic dipole operator utilised in this study may require supplementation with terms generated by two-body meson-exchange currents. Work in these directions is the focus of our current research. 

\section*{Acknowledgments}
This work was partially supported by STFC Grant Nos. ST/W005832/1, ST/P003885/1, ST/W50791X/1, and ST/V001035/1, and by a Leverhulme Trust Research Project Grant.
We acknowledge the CSC-IT Center for Science Ltd., Finland, and the IFT Computer Center of the University of Warsaw, Poland, for allocating computational resources. 
This project was partly undertaken on the Viking Cluster, a high-performance compute facility provided by the University of York. We are grateful for computational support from the University of York High Performance Computing service, Viking, and the Research Computing team.
We thank Grammarly\textsuperscript{\textregistered} for its support with English writing.

\bibliographystyle{iopart-num}
\bibliography{Paper_EM_moments,jacwit42}

\providecommand{\newblock}{}
\begin{thebibliography}{10}
\expandafter\ifx\csname url\endcsname\relax
  \def\url#1{{\tt #1}}\fi
\expandafter\ifx\csname urlprefix\endcsname\relax\def\urlprefix{URL }\fi
\providecommand{\eprint}[2][]{\url{#2}}

\bibitem{(Yan23)}
Yang X, Wang S, Wilkins S and Ruiz R~G 2023 {\em Progress in Particle and
  Nuclear Physics\/} {\bf 129} 104005 ISSN 0146-6410
  \urlprefix\url{https://www.sciencedirect.com/science/article/pii/S0146641022000631}

\bibitem{Castel1990}
Castel B and Towner I~S 1990 {\em Modern Theories of Nuclear Moments\/} (Oxford
  University PressOxford) ISBN 9780198517283
  \urlprefix\url{https://academic.oup.com/book/54719}

\bibitem{Neyens2003}
Neyens G 2003 {\em Reports on Progress in Physics\/} {\bf 66}(7) 1251--1251
  ISSN 0034-4885

\bibitem{(Sch37)}
Schmidt T 1937 {\em Z. Physik\/} {\bf 106} 358--361
  \urlprefix\url{https://doi.org/10.1007/BF01338744}

\bibitem{(Bon23c)}
Bonnard J, Dobaczewski J, Danneaux G and Kortelainen M 2023 {\em Physics
  Letters B\/} {\bf 843} 138014 ISSN 0370-2693
  \urlprefix\url{https://doi.org/10.1016/j.physletb.2023.138014}

\bibitem{Yordanov2020}
Yordanov D~T, Rodr{\'{i}}guez L~V, Balabanski D~L, Biero{\'{n}} J, Bissell M~L,
  Blaum K, Cheal B, Ekman J, Gaigalas G, {Garcia Ruiz} R~F, Georgiev G, Gins W,
  Godefroid M~R, Gorges C, Harman Z, Heylen H, J{\"{o}}nsson P, Kanellakopoulos
  A, Kaufmann S, Keitel C~H, Lagaki V, Lechner S, Maa{\ss} B,
  Malbrunot-Ettenauer S, Nazarewicz W, Neugart R, Neyens G,
  N{\"{o}}rtersh{\"{a}}user W, Oreshkina N~S, Papoulia A, Pyykk{\"{o}} P,
  Reinhard P~G, Sailer S, S{\'{a}}nchez R, Schiffmann S, Schmidt S, Wehner L,
  Wraith C, Xie L, Xu Z and Yang X 2020 {\em Commun. Phys.\/} {\bf 3} 107 ISSN
  2399-3650 \urlprefix\url{https://www.nature.com/articles/s42005-020-0348-9}

\bibitem{(Lec23)}
Lechner S, Miyagi T, Xu Z, Bissell M, Blaum K, Cheal B, Devlin C, {Garcia Ruiz}
  R, Ginges J, Heylen H, Holt J, Imgram P, Kanellakopoulos A, Koszor{\'u}s
  {\'A}, Malbrunot-Ettenauer S, Neugart R, Neyens G, N{\"o}rtersh{\"a}user W,
  Plattner P, Rodr{\'\i}guez L, Sanamyan G, Stroberg S, Utsuno Y, Yang X and
  Yordanov D 2023 {\em Physics Letters B\/} {\bf 847} 138278 ISSN 0370-2693
  \urlprefix\url{https://www.sciencedirect.com/science/article/pii/S0370269323006123}

\bibitem{(Var88)}
Varshalovich D, Moskalev A and Khersonskii V 1988 {\em Quantum Theory of
  Angular Momentum\/} (World Scientific, Singapore)

\bibitem{(Sas22c)}
Sassarini P~L, Dobaczewski J, Bonnard J and {Garcia Ruiz} R~F 2022 {\em Journal
  of Physics G: Nuclear and Particle Physics\/} {\bf 49} 11LT01
  \urlprefix\url{https://doi.org/10.1088/1361-6471/ac900a}

\bibitem{(Als23)}
Alswaidawi O~A and Alzubadi A~A 2023 {\em International Journal of Modern
  Physics E\/} {\bf 32} 2350020
  \urlprefix\url{https://doi.org/10.1142/S0218301323500209}

\bibitem{(Ye23)}
Ye H, Li J, Yang D, Jin H and Huang X 2023 {\em Communications in Theoretical
  Physics\/} {\bf 75} 025302
  \urlprefix\url{https://dx.doi.org/10.1088/1572-9494/aca07f}

\bibitem{(Pow22a)}
Powel R, Brown B~A, Holt J~D, Klose A, K\"onig K, Lantis J, Minamisono K,
  Miyagi T and Pineda S 2022 {\em Phys. Rev. C\/} {\bf 105}(3) 034310
  \urlprefix\url{https://link.aps.org/doi/10.1103/PhysRevC.105.034310}

\bibitem{(Ish23)}
Ishibashi Y, Gladkov A, Ichikawa Y, Takamine A, Nishibata H, Sato T, Yamazaki
  H, Abe T, Daugas J~M, Egami T, Fujita T, Georgiev G, Imamura K, Kawaguchi T,
  Kobayashi W, Nakamura Y, Ozawa A, Sanjo M, Shimizu N, Tominaga D, Tao L~C,
  Asahi K and Ueno H 2023 {\em Phys. Rev. C\/} {\bf 107}(2) 024306
  \urlprefix\url{https://link.aps.org/doi/10.1103/PhysRevC.107.024306}

\bibitem{(Sah24)}
Sahoo S, Srivastava P~C, Shimizu N and Utsuno Y 2024 {\em Phys. Rev. C\/} {\bf
  110}(2) 024306
  \urlprefix\url{https://link.aps.org/doi/10.1103/PhysRevC.110.024306}

\bibitem{(Bor24)}
Borzov I and Tolokonnikov S 2024 {\em Phys. Atom. Nuclei\/} {\bf 87} 423--435
  \urlprefix\url{https://doi.org/10.1134/S1063778824700376}

\bibitem{(Yue24)}
Yue Z, Andreyev A, Barzakh A, Borzov I, Cubiss J, Algora A, Au M, Balogh M,
  Bara S, Bark R, Bernerd C, Borge M, Brugnara D, Chrysalidis K, Cocolios T,
  {De Witte} H, Favier Z, Fraile L, Fynbo H, Gottardo A, Grzywacz R, Heinke R,
  Illana A, Jones P, Judson D, Korgul A, K{\"o}ster U, Labiche M, Le L, Lica R,
  Madurga M, Marginean N, Marsh B, Mihai C, N{\'a}cher E, Neacsu C, Nita C,
  Olaizola B, Orce J, Page C, Page R, Pakarinen J, Papadakis P, Penyazkov G,
  Perea A, Piersa-Si{\l}kowska M, Podoly{\'a}k Z, Prosnyak S, Reis E, Rothe S,
  Sedlak M, Skripnikov L, Sotty C, Stegemann S, Tengblad O, Tolokonnikov S,
  Ud{\'\i}as J, {Van Duppen} P, Warr N and Wojtaczka W 2024 {\em Physics
  Letters B\/} {\bf 849} 138452 ISSN 0370-2693
  \urlprefix\url{https://www.sciencedirect.com/science/article/pii/S037026932400011X}

\bibitem{(Tab23)}
Tabar E, Yakut H, Ho{\c{s}}g{\"o}r G and Kemah E 2023 {\em Eur. Phys. J. A\/}
  {\bf 59} 307 \urlprefix\url{https://doi.org/10.1140/epja/s10050-023-01223-0}

\bibitem{(Gav22)}
Gavrielov N, Leviatan A and Iachello F 2022 {\em Phys. Rev. C\/} {\bf 105}(1)
  014305 \urlprefix\url{https://link.aps.org/doi/10.1103/PhysRevC.105.014305}

\bibitem{(Gav23)}
Gavrielov N 2023 {\em Phys. Rev. C\/} {\bf 108}(1) 014320
  \urlprefix\url{https://link.aps.org/doi/10.1103/PhysRevC.108.014320}

\bibitem{(Cub23)}
Cubiss J~G, Andreyev A~N, Barzakh A~E, Van~Duppen P, Hilaire S, P\'eru S,
  Goriely S, Al~Monthery M, Althubiti N~A, Andel B, Antalic S, Atanasov D,
  Blaum K, Cocolios T~E, Day~Goodacre T, de~Roubin A, Farooq-Smith G~J, Fedorov
  D~V, Fedosseev V~N, Fink D~A, Gaffney L~P, Ghys L, Harding R~D, Huyse M, Imai
  N, Joss D~T, Kreim S, Lunney D, Lynch K~M, Manea V, Marsh B~A,
  Martinez~Palenzuela Y, Molkanov P~L, Neidherr D, O'Neill G~G, Page R~D,
  Prosnyak S~D, Rosenbusch M, Rossel R~E, Rothe S, Schweikhard L, Seliverstov
  M~D, Sels S, Skripnikov L~V, Stott A, Van~Beveren C, Verstraelen E, Welker A,
  Wienholtz F, Wolf R~N and Zuber K 2023 {\em Phys. Rev. Lett.\/} {\bf 131}(20)
  202501
  \urlprefix\url{https://link.aps.org/doi/10.1103/PhysRevLett.131.202501}

\bibitem{(Rys22a)}
Ryssens W, Scamps G, Goriely S and Bender M 2022 {\em Eur. Phys. J. A\/} {\bf
  58} 246 \urlprefix\url{https://doi.org/10.1140/epja/s10050-022-00894-5}

\bibitem{(Web23)}
Weber F, Albrecht-Sch\"onzart T~E, Allehabi S~O, Berndt S, Block M, Dorrer H,
  D\"ullmann C~E, Dzuba V~A, Ezold J~G, Flambaum V~V, Gadelshin V, Goriely S,
  Harvey A, Heinke R, Hilaire S, Kaja M, Kieck T, Kneip N, K\"oster U, Lantis
  J, Mokry C, M\"unzberg D, Nothhelfer S, Oberstedt S, P\'eru S, Raeder S,
  Runke J, Sonnenschein V, Stemmler M, Studer D, Th\"orle-Pospiech P, Tomita H,
  Trautmann N, Van~Cleve S, Warbinek J and Wendt K 2023 {\em Phys. Rev. C\/}
  {\bf 107}(3) 034313
  \urlprefix\url{https://link.aps.org/doi/10.1103/PhysRevC.107.034313}

\bibitem{(Mor22)}
Moreno O, Sarriguren P, Algora A, Fraile L~M and Orrigo S~E~A 2022 {\em Phys.
  Rev. C\/} {\bf 106}(3) 034317
  \urlprefix\url{https://link.aps.org/doi/10.1103/PhysRevC.106.034317}

\bibitem{(Zho25)}
Zhou E, Ding C, Yao J, Bally B, Hergert H, Jiao C and Rodríguez T 2025 {\em
  Eur. Phys. J. A\/} {\bf 865} 139464
  \urlprefix\url{https://doi.org/10.1016/j.physletb.2025.139464}

\bibitem{(Fro22)}
Frosini M, Duguet T, Ebran J, Bally B, Mongelli T, Rodríguez T, Roth R and
  Somà V 2022 {\em Eur. Phys. J. A\/} {\bf 58} 63
  \urlprefix\url{https://link.springer.com/article/10.1140/epja/s10050-022-00693-y}

\bibitem{(Sar23)}
Sarma C and Srivastava P~C 2023 {\em Journal of Physics G: Nuclear and Particle
  Physics\/} {\bf 50} 045105
  \urlprefix\url{https://dx.doi.org/10.1088/1361-6471/acb962}

\bibitem{(Chu23)}
Choudhary P and Srivastava P 2023 {\em Eur. Phys. J. A\/} {\bf 59} 97
  \urlprefix\url{https://doi.org/10.1140/epja/s10050-023-01013-8}

\bibitem{(Mul24)}
M{\"u}ller P, Kaufmann S, Miyagi T, Billowes J, Bissell M, Blaum K, Cheal B,
  {Garcia Ruiz} R, Gins W, Gorges C, Heylen H, Kanellakopoulos A,
  Malbrunot-Ettenauer S, Neugart R, Neyens G, N{\"o}rtersh{\"a}user W,
  Ratajczyk T, Rodr{\'\i}­guez L, S{\'a}nchez R, Sailer S, Schwenk A, Wehner
  L, Wraith C, Xie L, Xu Z, Yang X and Yordanov D 2024 {\em Physics Letters
  B\/} {\bf 854} 138737 ISSN 0370-2693
  \urlprefix\url{https://www.sciencedirect.com/science/article/pii/S0370269324002958}

\bibitem{(Zho24)}
Zhou E~F, Wu X~Y and Yao J~M 2024 {\em Phys. Rev. C\/} {\bf 109}(3) 034305
  \urlprefix\url{https://link.aps.org/doi/10.1103/PhysRevC.109.034305}

\bibitem{(Bal23)}
Bally B, Giacalone G and Bender M 2023 {\em Eur. Phys. J. A\/} {\bf 59} 58
  \urlprefix\url{https://doi.org/10.1140/epja/s10050-023-00955-3}

\bibitem{(Iva22)}
Ivanova D, Lalkovski S, Costache C, Kisyov S, Mihai C, M\ifmmode~\check{a}\else
  \v{a}\fi{}rginean N, Petkov P, Atanasova L, Bucurescu D, Cakirli R~B,
  Carpenter M~P, Casten R~F, C\ifmmode \check{a}\else~\v{a}\fi{}ta Danil G,
  C\ifmmode \check{a}\else~\v{a}\fi{}ta Danil I, Deleanu D, Filipescu D, Florea
  N, Gheorghe I, Ghi\ifmmode \mbox{\c{t}}\else
  \c{t}\fi{}\ifmmode~\check{a}\else \v{a}\fi{} D, Glodariu T, Kondev F~G, Lica
  R, M\ifmmode~\check{a}\else \v{a}\fi{}rginean R, Negret A, Pascu S, Sava T,
  Stefanova E~A, Stroe L, Suliman G, Suvaila R, Yordanov O and Zamfir N~V 2022
  {\em Phys. Rev. C\/} {\bf 105}(3) 034337
  \urlprefix\url{https://link.aps.org/doi/10.1103/PhysRevC.105.034337}

\bibitem{(deG22)}
{de Groote} R, Moreno J, Dobaczewski J, Koszor\'us A, Moore I, Reponen M, Sahoo
  B and Yuan C 2022 {\em Physics Letters B\/} {\bf 827} 136930 ISSN 0370-2693
  \urlprefix\url{https://www.sciencedirect.com/science/article/pii/S0370269322000648}

\bibitem{(Ver22b)}
Vernon A~R, {Garcia Ruiz} R~F, Miyagi T, Binnersley C~L, Billowes J, Bissell
  M~L, Bonnard J, Cocolios T~E, Dobaczewski J, Farooq-Smith G~J, Flanagan K~T,
  Georgiev G, Gins W, de~Groote R~P, Heinke R, Holt J~D, Hustings J, Koszor\'us
  {\'A}, Leimbach D, Lynch K~M, Neyens G, Stroberg S~R, Wilkins S~G, Yang X~F
  and Yordanov D~T 2022 {\em Nature\/} {\bf 607} 260
  \urlprefix\url{https://doi.org/10.1038/s41586-022-04818-7}

\bibitem{(Nie23)}
Nies L, Atanasov D, Athanasakis-Kaklamanakis M, Au M, Blaum K, Dobaczewski J,
  Hu B~S, Holt J~D, Karthein J, Kulikov I, Litvinov Y~A, Lunney D, Manea V,
  Miyagi T, Mougeot M, Schweikhard L, Schwenk A, Sieja K and Wienholtz F 2023
  {\em Phys. Rev. Lett.\/} {\bf 131}(2) 022502
  \urlprefix\url{https://link.aps.org/doi/10.1103/PhysRevLett.131.022502}

\bibitem{(Gra23)}
Gray T, Stuchbery A, Dobaczewski J, Blazhev A, Alshammari H, Bignell L, Bonnard
  J, Coombes B, Dowie J, Gerathy M, Kib{\'e}di T, Lane G, McCormick B, Mitchell
  A, Nicholls C, Pope J, Reinhard P~G, Spinks N and Zhong Y 2023 {\em Physics
  Letters B\/} {\bf 847} 138268 ISSN 0370-2693
  \urlprefix\url{https://www.sciencedirect.com/science/article/pii/S0370269323006020}

\bibitem{(deG24)}
{de Groote} R, Nesterenko D, Kankainen A, Bissell M, Beliuskina O, Bonnard J,
  Campbell P, Canete L, Cheal B, Delafosse C, {de Roubin} A, Devlin C,
  Dobaczewski J, Eronen T, {Garcia Ruiz} R, Geldhof S, Gins W, Hukkanen M,
  Imgram P, Mathieson R, Koszor{\'u}s {\'A}, Moore I, Pohjalainen I, Reponen M,
  {van den Borne} B, Vil{\'e}n M and Zadvornaya S 2024 {\em Physics Letters
  B\/} {\bf 848} 138352 ISSN 0370-2693
  \urlprefix\url{https://www.sciencedirect.com/science/article/pii/S037026932300686X}

\bibitem{(Kar24)}
Karthein J, Ricketts C~M, Garcia~Ruiz R~F, Billowes J, Binnersley C~L, Cocolios
  T~E, Dobaczewski J, Farooq-Smith G~J, Flanagan K~T, Georgiev G, Gins W,
  de~Groote R~P, Gustafsson F~P, Holt J~D, Kanellakopoulos A, Koszor{\'u}s
  {\'A}, Leimbach D, Lynch K~M, Miyagi T, Nazarewicz W, Neyens G, Reinhard P~G,
  Sahoo B~K, Vernon A~R, Wilkins S~G, Yang X~F and Yordanov D~T 2024 {\em Nat.
  Phys.\/} {\bf 20} 1719–1725
  \urlprefix\url{https://doi.org/10.1038/s41567-024-02612-y}

\bibitem{(Dob09g)}
Dobaczewski J, Satu{\l}a W, Carlsson B, Engel J, Olbratowski P, Powa{\l}owski
  P, Sadziak M, Sarich J, Schunck N, Staszczak A, Stoitsov M, Zalewski M and
  Zdu{\'n}czuk H 2009 {\em Comput. Phys. Commun.\/} {\bf 180} 2361 -- 2391 ISSN
  0010-4655
  \urlprefix\url{http://www.sciencedirect.com/science/article/pii/S0010465509002598}

\bibitem{(Rin80)}
Ring P and Schuck P 1980 {\em The Nuclear Many-Body Problem\/}
  (Springer-Verlag, Berlin)

\bibitem{(Dob04d)}
Dobaczewski J and Olbratowski P 2004 {\em Comput. Phys. Comm.\/} {\bf 158} 158
  -- 191 ISSN 0010-4655
  \urlprefix\url{http://www.sciencedirect.com/science/article/pii/S0010465504000797}

\bibitem{Sheikh_2021}
Sheikh J~A, Dobaczewski J, Ring P, Robledo L~M and Yannouleas C 2021 {\em
  Journal of Physics G: Nuclear and Particle Physics\/} {\bf 48} 123001
  \urlprefix\url{https://dx.doi.org/10.1088/1361-6471/ac288a}

\bibitem{(Kor12b)}
Kortelainen M, McDonnell J, Nazarewicz W, Reinhard P~G, Sarich J, Schunck N,
  Stoitsov M~V and Wild S~M 2012 {\em Phys. Rev. C\/} {\bf 85}(2) 024304
  \urlprefix\url{https://link.aps.org/doi/10.1103/PhysRevC.85.024304}

\bibitem{(Dob95b)}
Dobaczewski J and Dudek J 1995 {\em Phys. Rev. C\/} {\bf 52}(4) 1827--1839
  \urlprefix\url{https://link.aps.org/doi/10.1103/PhysRevC.52.1827}

\bibitem{(Per04c)}
Perli\'{n}ska E, Rohozi\'{n}ski S~G, Dobaczewski J and Nazarewicz W 2004 {\em
  Phys. Rev. C\/} {\bf 69}(1) 014316
  \urlprefix\url{http://link.aps.org/doi/10.1103/PhysRevC.69.014316}

\bibitem{(Ben02d)}
Bender M, Dobaczewski J, Engel J and Nazarewicz W 2002 {\em Phys. Rev. C\/}
  {\bf 65}(5) 054322
  \urlprefix\url{http://link.aps.org/doi/10.1103/PhysRevC.65.054322}

\bibitem{(Hel12)}
Hellemans V, Heenen P~H and Bender M 2012 {\em Phys. Rev. C\/} {\bf 85}(1)
  014326 \urlprefix\url{https://link.aps.org/doi/10.1103/PhysRevC.85.014326}

\bibitem{(Bender00)}
Bender M, Rutz K, Reinhard P~G and Maruhn J 2000 {\em Eur. Phys. J. A\/} {\bf
  8} 59
  \urlprefix\url{https://link.springer.com/article/10.1007/s10050-000-4504-z}

\bibitem{Dobaczewski_2021}
Dobaczewski J, Bączyk P, Becker P, Bender M, Bennaceur K, Bonnard J, Gao Y,
  Idini A, Konieczka M, Kortelainen M, Próchniak L, Romero A~M, Satuła W, Shi
  Y, Werner T~R and Yu L~F 2021 {\em Journal of Physics G: Nuclear and Particle
  Physics\/} {\bf 48} 102001
  \urlprefix\url{https://dx.doi.org/10.1088/1361-6471/ac0a82}

\bibitem{(Dob25a)}
{Dobaczewski J {\it et al}} 2025 Code {\sc hfodd}, version to be published

\bibitem{supp-SnGd}
{See Supplemental Material at [URL will be inserted by publisher] for tables of
  detailed results in numerical form. }

\bibitem{Stone2014}
Stone N~J 2014 {Table of Nuclear Magnetic Dipole and Electric Quadrupole
  Moments} Tech. rep. INDC International Nuclear Data Committee
  \urlprefix\url{https://www-nds.iaea.org/publications/indc/indc-nds-0658.pdf}

\bibitem{Stone2016}
Stone N 2016 {\em At. Data Nucl. Data Tables\/} {\bf 111-112} 1--28 ISSN
  0092640X

\bibitem{Stone2021}
Stone N 2021 {Table of nuclear electric quadrupole moments} Tech. rep. INDC
  International Nuclear Data Committee

\bibitem{Alkhazov1992}
Alkhazov G~D, Barzakh A~E, Huhnermann H, Kesper K, Mazumdar A, Moller W, Otto
  R, Pantelejev V~N, Poljakov A~G, Reese C and Wagner H 1992 {\em J. Phys. B
  At. Mol. Opt. Phys.\/} {\bf 25} 571--576 ISSN 0953-4075
  \urlprefix\url{https://iopscience.iop.org/article/10.1088/0953-4075/25/2/023}

\bibitem{Haiduke2006}
Haiduke R~L~A, da~Silva A~B~F and Visscher L 2006 {\em J. Chem. Phys.\/} {\bf
  125} ISSN 0021-9606
  \urlprefix\url{https://pubs.aip.org/jcp/article/125/6/064301/906644/The-nuclear-electric-quadrupole-moment-of-antimony}

\bibitem{(Lec21a)}
Lechner S, Xu Z~Y, Bissell M~L, Blaum K, Cheal B, De~Gregorio G, Devlin C~S,
  Garcia~Ruiz R~F, Gargano A, Heylen H, Imgram P, Kanellakopoulos A, Koszor\'us
  A, Malbrunot-Ettenauer S, Neugart R, Neyens G, N\"ortersh\"auser W, Plattner
  P, Rodr\'{\i}guez L~V, Yang X~F and Yordanov D~T 2021 {\em Phys. Rev. C\/}
  {\bf 104}(1) 014302
  \urlprefix\url{https://link.aps.org/doi/10.1103/PhysRevC.104.014302}

\bibitem{Gray}
{As discussed in~\protect\cite{(Gra23)}, deviations obtained in light Sn
  isotopes can be corrected using the pairing strength lowered by another 20\%
  }

\bibitem{Stone2019}
Stone N~J 2019 {Table of Recommended Nuclear Magnetic Dipole Moments: Part I -
  Long-lived States} Tech. rep. INDC International Nuclear Data Committee
  \urlprefix\url{https://nds.iaea.org/publications/indc/indc-nds-0794.pdf}

\bibitem{Stone2020}
Stone N~J 2020 {Table of Recommended Nuclear Magnetic Dipole Moments: Part II -
  Short-lived States} Tech. rep. INDC International Nuclear Data Committee
  \urlprefix\url{https://www-nds.iaea.org/publications/indc/indc-nds-0816.pdf}

\bibitem{(Dob97c)}
Dobaczewski J and Dudek J 1997 {\em Comput. Phys. Comm.\/} {\bf 102} 183 -- 209
  ISSN 0010-4655
  \urlprefix\url{http://www.sciencedirect.com/science/article/pii/S0010465597000052}

\bibitem{ENSDF.230918}
From {ENSDF} database as of september 18, 2023. {V}ersion available at
  \url{http://www.nndc.bnl.gov/ensarchivals/}

\bibitem{(Wan21)}
Wang M, Huang W, Kondev F, Audi G and Naimi S 2021 {\em Chinese Physics C\/}
  {\bf 45} 030003 \urlprefix\url{https://dx.doi.org/10.1088/1674-1137/abddaf}

\bibitem{(Pad65)}
Padgett D~W, Frank W~M and Brennan J~G 1965 {\em Nuclear Physics\/} {\bf 73}
  424--444 ISSN 0029-5582
  \urlprefix\url{https://www.sciencedirect.com/science/article/pii/0029558265906887}

\bibitem{(Che69)}
Chemtob M 1969 {\em Nuclear Physics A\/} {\bf 123} 449--470 ISSN 0375-9474
  \urlprefix\url{https://www.sciencedirect.com/science/article/pii/0375947469905132}

\bibitem{(Miy24)}
Miyagi T, Cao X, Seutin R, Bacca S, Ruiz R~F~G, Hebeler K, Holt J~D and Schwenk
  A 2024 {\em Phys. Rev. Lett.\/} {\bf 132}(23) 232503
  \urlprefix\url{https://link.aps.org/doi/10.1103/PhysRevLett.132.232503}

\end{thebibliography}

\end{document}